\documentclass[aps,showpacs,preprintnumbers,amsmath,amssymb]{revtex4}
\UseRawInputEncoding
\usepackage{dcolumn}
\usepackage{multirow}

\usepackage{float}
\usepackage{color}
\usepackage{graphicx}
\usepackage{epstopdf}

\def\mnras{MNRAS}

\def\actaa{Acta Astron}
\def\aap{Astron Astroph}

\def\scpma{Sci. China Phys. Mech. Astron.}

\makeatletter

\def \be {\begin{equation}}
\def \ee {\end{equation}}
\def \bea {\begin{eqnarray}}
\def \eea {\end{eqnarray}}

\begin{document}
\baselineskip=0.8 cm
\title{\bf Geometrically thick equilibrium tori around a dyonic black hole with quasi-topological electromagnetism }
\author{Xuan Zhou$^{1}$,
    Songbai Chen$^{1,2}$\footnote{Corresponding author: csb3752@hunnu.edu.cn},
    Jiliang Jing$^{1,2}$ \footnote{jljing@hunnu.edu.cn}}
    \affiliation{$^1$Department of Physics,  Key Laboratory of Low Dimensional Quantum Structures
    and Quantum Control of Ministry of Education, Institute of Interdisciplinary Studies, Synergetic Innovation Center for Quantum Effects and Applications, Hunan
    Normal University,  Changsha, Hunan 410081, People's Republic of China
    \\
    $ ^2$Center for Gravitation and Cosmology, College of Physical Science and Technology, Yangzhou University, Yangzhou 225009, People's Republic of China}

\begin{abstract}
\baselineskip=0.6 cm
\begin{center}
{\bf Abstract}
\end{center}

We study geometrically thick and non-self gravitating equilibrium tori orbiting a static spherically symmetric dyonic black hole with quasi-topological electromagnetism. Our results show that the electric and magnetic charges together with the coupling parameter in the quasi-topological electromagnetism lead to a much richer class of equilibrium tori. There are a range of parameters which allow for the existence of double tori. The properties of double equilibrium tori become far richer. There exist transitions between single torus and double tori solutions as we change the specific angular momentum of the fluid. These properties of equilibrium tori could help to understand the dyonic black hole and its thick accretion disk.
\end{abstract}

\maketitle
\newpage

\section{Introduction}

The presence of the bright zones in the images of the supermassive black
holes M87* \cite{EHT1,EHT2,EHT3,EHT4,EHT5,EHT6} and Sgr A* \cite{EHT7} means that a black hole at the centre of a galaxy must be surrounded by an accretion disk. This is because conversion of gravitational energy into heat and radiation in the matter accretion  is the most efficient so that
the strong electromagnetic radiation emitted by  disk can illuminate regions near black holes.
In the real astrophysical systems, the matter accretion is a highly complicated dynamic process and its complete description must resort to high precise numerical calculations, such as general relativistic
magnetohydrodynamics (GRMHD) simulations \cite{Fishbone:1976lv,EventHorizonTelescope:2019pcy,Font:2008fka,Roy:2015kma,Porth:2016rfi,Anton:2005gi}. However, in the past few decades, a simple model of geometrically thick and stationary tori orbiting black holes
has been attracted a lot of attention. In this model,  matter is assumed to be stationarily rotating and without actually approaching the black hole \cite{1974AcA2445A,1971AcA2181A,1978A&A63209K,1978AA63221A,1980AcA301J,1980A&A8823P,1980ApJ242772A,
1982MitAG5727P,1982ApJ253897P}. Moreover, the self-gravity of the
fluid body is neglected. Interestingly, such stationarily rotating perfect fluid tori, known as Polish doughnuts,  are exact analytical solutions of the relativistic Euler equation \cite{Chakraborty,Chakraborty2,liux}. Due to the fluid being in equilibrium, Polish doughnuts are often
used as an initial condition for numerical simulations of accretion flows. Additionally, the features and configurations of these geometrically thick equilibrium tori carry a lot of important information on the spacetime in strong field regions because the motion of fluid is very close to the event horizon of black holes. Thus, study of geometrically thick equilibrium tori can offer a potential way of probing
spacetime characteristic properties imprinted on the equilibrium tori.

The geometrically thick non-self gravitating equilibrium tori orbiting black holes have been studied for many spacetimes in
general relativity and in other alternative theories of gravity \cite{Komissarov:2004qu,2006MNRAS3683K,Font:2002ci,Daigne:2003tf,Font:2002bi,PMID:28179840,Liu:2022cph,2022EPJC82190K}.
Recently, equilibrium tori have been investigated for a spherically symmetric black hole in Born-Infeld teleparallel gravity, which show that there is only a single torus as in the Schwarzschild spacetime, but the teleparallel gravity parameter leads to that the size of the torus becomes small \cite{Bahamonde:2021srr}. The parameterised Rezzolla-Zhidenko black hole is found to have a much richer class of equilibrium tori \cite{2015MNRAS52222M}. There exist standard single-torus and non-standard double-tori solutions within the allowed range of parameters and the transitions between single torus and double tori solutions can occur by regulating the specific angular momentum of the fluid.
Moreover, the magnetized equilibrium tori around Kerr black hole with scalar hair have been studied with the constant
angular momentum model \cite{2019PhRvD99d3002G} and the nonconstant angular momentum model \cite{2021PhRvD104j3008G}. The properties of the stationary thick tori differed from that in the Kerr case \cite{2006MNRAS3683K} could be used to further constrain the no-hair hypothesis with future observations. The geometrically thick equilibrium tori with the constant angular momentum model are investigated for a static spherically symmetric black
hole in $f(R)$-gravity with a Yukawa-like modification\cite{2021PhRvD103l4009C}. It is shown that configurations of the tori own the notable differences from those in the usual black hole in the general relativity. Moreover, the magnetised equilibrium tori around a  Kerr black hole are also studied with the non-constant specific angular momentum distribution model \cite{Gimeno-Soler:2023anr}. The configurations of the geometrically thick tori have been also studied in the background of compact object with a quadrupole moment \cite{qq1} and of a binary black hole system \cite{qq2}.

Recently, a dyonic black hole solution is obtained in the frame of a quasi-topological electromagnetism \cite{2020scpma63L}, which is a higher-order extension with the bilinear norm of Maxwell theory. Although under some appropriate conditions,
quasi-topological terms have no contribution to Maxwell equation and energy-momentum tensor, they can nontrivially modify the dyonic
solutions. The dyonic black hole solution carries both electric and magnetic charges. Although magnetic charges  have not been observed in nature, they can arise as topologically nontrivial classical solution in some spontaneously broken gauge theories. It looks very difficult to generate artificially
 dyonic black holes because it would involve producing or gathering lots of magnetic monopoles and then collapsing them into a black hole.
One plausible mechanism is to produce first a large number of monopoles and anti-monopoles in the early universe, then at larger scales one needs large primordial fluctuations that produce black holes \cite{2020jhep79}.
The dyonic black hole solution \cite{2020scpma63L} in quasi-topological electromagnetism  exhibits some unusual properties. In certain parameter region, the black hole
solution possess four horizons and three photon spheres. The energy condition is analyzed for the existence of three photon spheres in the dyonic
black hole \cite{2023prd10712G}. Especially, this dyonic black hole solution can be used to construct Dyson-like
spheres around the black hole \cite{2023prddw}, at which a massive particle remains at rest
with respect to a static asymptotic static observer. These studies shed new light on understanding the dyonic black hole solution with
quasi-topological electromagnetism. The main motivation of this paper is to study the geometrically thick equilibrium tori around the dyonic black hole with quasi-topological electromagnetism and to see what new properties of the equilibrium tori in this case.

The paper is organized as follows: In Sec. II, we briefly introduce the dyonic black hole solution and Polish doughnut model of thick accretion disks. In Sec.III, we present properties of equilibrium tori around  the dyonic black hole. Finally, we present a summary.

\section{Dyonic black hole and Polish doughnut model of thick accretion disks}

\begin{figure}[ht]
\includegraphics[width=3.8cm ]{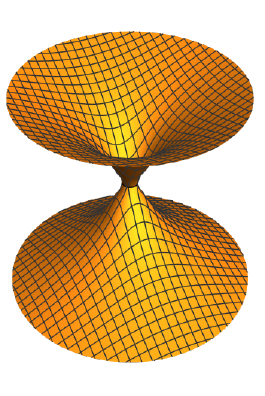}\;\;\;\;\includegraphics[width=3.8cm ]{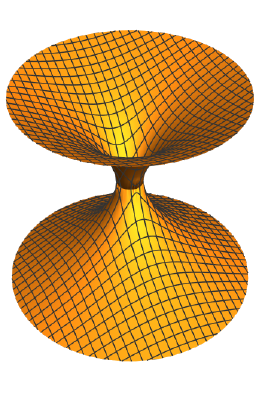}\;\;\;\;
\includegraphics[width=3.8cm ]{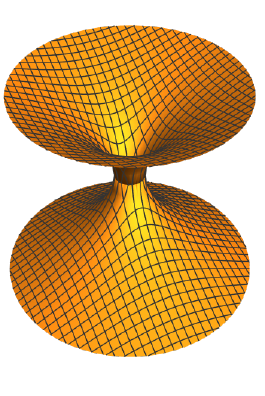}
\caption{The embedding diagrams of the metric (\ref{metric0}). The left, middle and right panels correspond to the cases $\alpha=6$, $9$ and $12$, respectively. Here we set $M=1$, $p=0.14$, and $q=1.1$. }\label{diagram}
\end{figure}

Let us now to briefly review the dyonic black hole \cite{2020scpma63L}, which is a static spherically symmetric black hole solution in the quasi-topological electromagnetism. The action has a form \cite{2020scpma63L}
\begin{equation}
S=\frac{1}{16 \pi} \int \sqrt{-g} d^4 x\left(R-\alpha_1 F^2-\alpha_2\left(\left(F^2\right)^2-2 F^{(4)}\right)\right), \label{eq12}
\end{equation}
where the field strength is $F^2 =-F_\nu^\mu F_\mu^\nu$ and $F^{(4)}=F_\nu^\mu F_\rho^\nu F_\sigma^\rho F_\mu^\sigma$. The Maxwell field strength reads $F_{\mu\nu}=\partial_{\nu}A_{\mu}-\partial_{\mu}A_{\nu}$ with $A_{\mu}$
the vector potential. The coupling parameters $\alpha_1$ and $\alpha_2$ are two non-negative constants, which respectively correspond to the standard Maxwell and quasi-topological electromagnetic actions.
 The action (\ref{eq12}) admits a spherically symmetric dyonic black hole solution with a metric form \cite{2020scpma63L}
\begin{equation}
d s^2=-f(r) d t^2+\frac{1}{f(r)} d r^2+r^2\left(d \theta^2+\sin ^2 \theta d \phi^2\right),\label{metric0}
\end{equation}
with
\begin{equation}
f(r)=1-\frac{2 M}{r}+\frac{\alpha_1 \tilde{p}^2}{r^2}+\frac{\tilde{q}^2}{\alpha_1 r^2}{ }_2 F_1\left(\frac{1}{4}, 1 ; \frac{5}{4} ;-\frac{4  \tilde{p}^2 \alpha_2}{r^4 \alpha_1}\right),
\end{equation}
where $_2 F_1(a,b,c; x)$ is the Hypergeometric function. This solution owns three integration constants $M$, $\tilde{q}$ and $\tilde{p}$, which are respectively related to the mass, electric and magnetic charges.
The corresponding ansatz of the Maxwell field is
 \begin{equation}
   A_\mu d x^\mu = \phi \left(r\right) d t+\tilde{p} \cos \theta d \phi,\label{ev-au}
 \end{equation}
 with
 \begin{equation}
   \phi^{'}\left(r\right)= -\frac{\tilde{q} r^2}{\alpha_1 r^4 + 4\alpha_2 \tilde{p}^2}.
 \end{equation}
The electric and magnetic charges are given by \cite{2020scpma63L}
\begin{equation}
  Q_e =\frac{1}{4\pi}\int \tilde{F}^{01} =\tilde {q},\quad\quad\quad Q_m=\frac{1}{4\pi \alpha_1}\int F= \frac{\tilde{p}}{\alpha_1},
 \end{equation}
 where
 \begin{equation}
  \tilde{F}^{\mu\nu}=4\alpha_1F^{\mu\nu}+8\alpha_2(F^2F^{\mu\nu}-2F^{\mu\rho}F^{\sigma}_{\;\rho}F^{\nu}_{\;\sigma}).
 \end{equation}
It is obvious that the electric and magnetic charges enter the metric asymmetrically and
then the electromagnetic duality breaks down by the factor $\alpha_2$.
From Eq.(\ref{ev-au}), we find that there exist only the radial electric field and the radial magnetic field. The corresponding electric intensity $E$ and magnetic induction intensity $B$ in the local orthonormal frame can be expressed as \cite{Zhang:2022klr}
 \begin{equation}
  E=\frac{\tilde{q} r^2}{\alpha_1 r^4 + 4\alpha_2 \tilde{p}^2},\quad\quad \quad B=\frac{\tilde{p}}{r^2}.\label{elemag}
 \end{equation}
\begin{figure}[ht]
\includegraphics[width=5.0cm ]{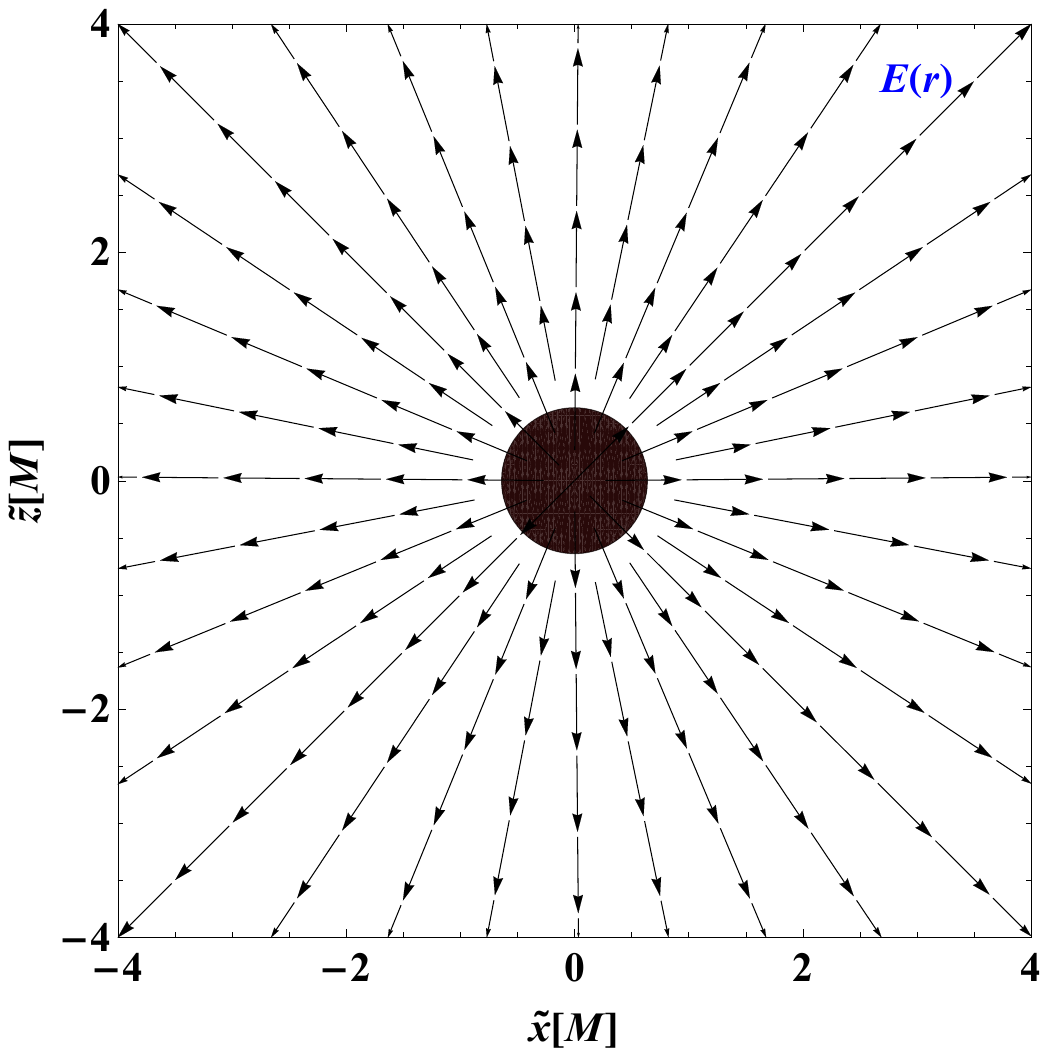}\includegraphics[width=5.0cm ]{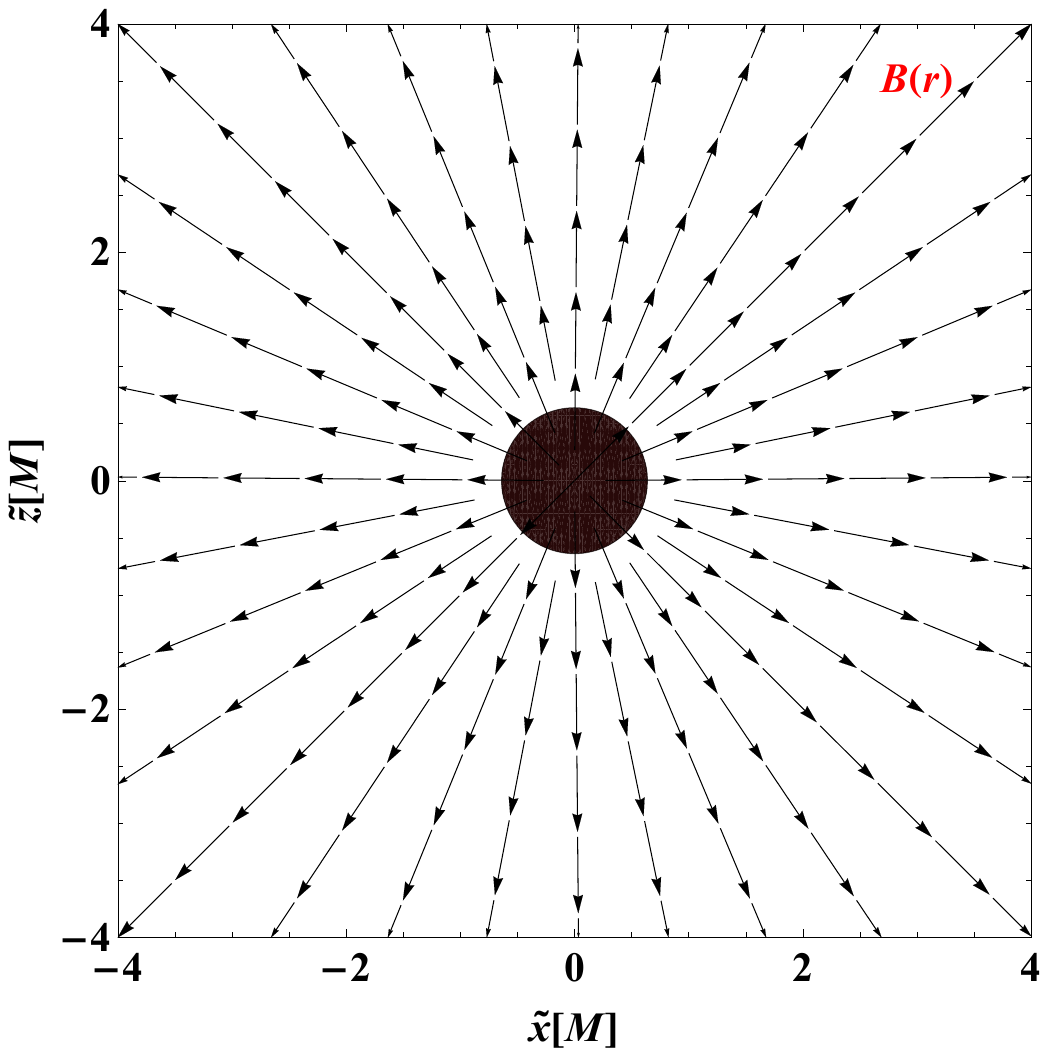}
\includegraphics[width=5.0cm ]{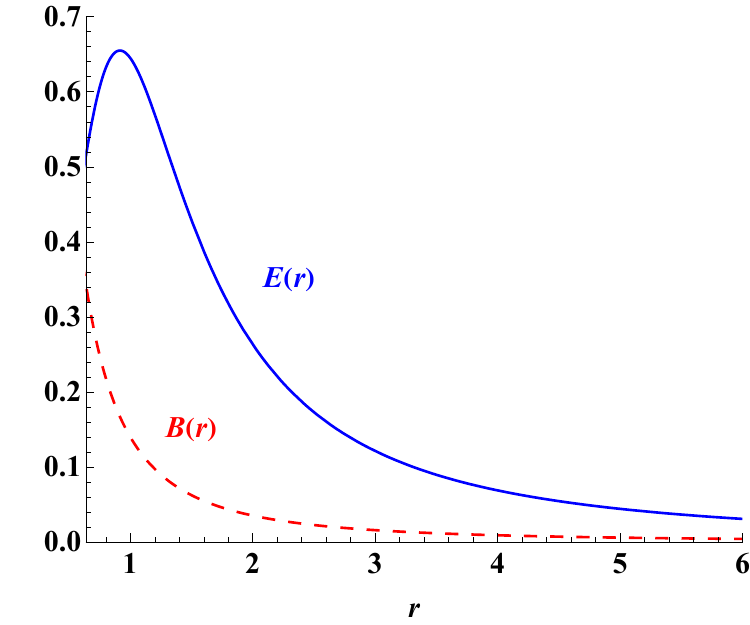}
\caption{Distributions of electric and magnetic fields for the dyonic  black hole with an specific set of parameters $\tilde{p}=0.14$, $\tilde{q}=1.1$, $\alpha_1=1$ and $\alpha_2 =9$. }\label{f1em}
\end{figure}
Fig.(\ref{f1em}) shows distributions of electric and magnetic fields for the dyonic  black hole with an specific set of parameters $\tilde{p}=0.14$, $\tilde{q}=1.1$, $\alpha_1=1$ and $\alpha_2 =9$.
The magnetic induction intensity monotonically decreases with the distance away from the central black hole, while the electric field intensity first increases and then decreases due to the effect of the magnetic charge parameter.
Moreover, the electric and magnetic field energy can be expressed as \cite{2020scpma63L}
\begin{equation}
 \mathcal{H}=\alpha_1(E^2+B^2)+4\alpha_2(EB)^2=\frac{\tilde{q}^2}{\alpha_1 r^4 + 4\alpha_2 \tilde{p}^2}+\frac{\alpha_1\tilde{p}^2}{r^4}
 \end{equation}
Since the energy $ \mathcal{H}$ is nonnegative in the case $\tilde{p}=0$,  the parameter $\alpha_1$ is nonnegative. However, when $\alpha_1=0$, the electric potential $\phi(r)=\phi_0-\frac{\tilde{q}r^3}{12\alpha_2 \tilde{p}^2}$ is divergent at spatial infinite. Thus, the parameter $\alpha_1$ is strictly positive. The metric function $f(r)$ can be expanded as
\begin{equation}
f(r)=1-\frac{2 M}{r}+\frac{\alpha_1 \tilde{p}^2}{r^2}+\frac{\tilde{q}^2}{\alpha_1 r^2}+\alpha_2\mathcal{F}(r,\tilde{p},\tilde{q},\alpha_1 ,\alpha_2),
 \end{equation}
where $\mathcal{F}(r,\tilde{p},\tilde{q},\alpha_1 ,\alpha_2)$ is a non-zero function. The function $f(r)$ can be reduce to that in the Schwarzschild spacetime only if $\alpha_2=0$ and $\alpha_1=\pm i\tilde{q}/\tilde{p}$. Thus, for the real non-zero value of $\alpha_1$,  the solution (\ref{metric0}) does not degenerate to the Schwarzschild one. For the convenience, we rescale the parameters as
 $q=\tilde{q}/\sqrt{\alpha_1}$, $p=\tilde{p}\sqrt{\alpha_1}$, and $\alpha=\alpha_2/\alpha^2_1$, which makes the metric function $f(r)$ more simple, i.e., $
f(r)=1-\frac{2 M}{r}+\frac{p^2}{r^2}+\frac{q^2}{ r^2}{ }_2 F_1(\frac{1}{4}, 1 ; \frac{5}{4} ;-\frac{4  p^2 \alpha}{r^4 })$.
To visualize the spacetime (\ref{metric0}), we present the embedding diagrams in Fig.\ref{diagram} where the  equatorial slice $\theta=\pi/2$ at some fix moment in time $t=$constant is embed into three-dimensional Euclidean
space $ds^2=dz^2+dr^2+r^2d\phi^2$ with $dz=\pm \sqrt{1/f-1}dr$.
Moreover, the metric (\ref{metric0}) is asymptotically flat and satisfies the dominant energy condition \cite{2020scpma63L}.
Comparing with the usual Reissner-Nordstr\"{o}m black hole, the black hole (\ref{metric0}) could have some interesting spacetime structures, for example,
there are four black hole horizons and three photon spheres in certain parameter regions. Moreover, in the case of pure electric charge (i.e., $p=0$) or pure magnetic charge (i.e., $q=0$), the metric of the dyonic black hole solution (\ref{metric0}) can the Reissner-Nordstr\"{o}m metric with pure electric charge or pure magnetic charge. However, when electric and magnetic charges emerge simultaneously,  the dyonic black hole solution (\ref{metric0}) differs from the Reissner-Nordstr\"{o}m solution in general relativity. Thus, the further studies of  the dyonic black hole solution (\ref{metric0}) could deepen the understanding gravity and quasi-topological electromagnetism.
\begin{figure}[ht]
\includegraphics[width=5cm ]{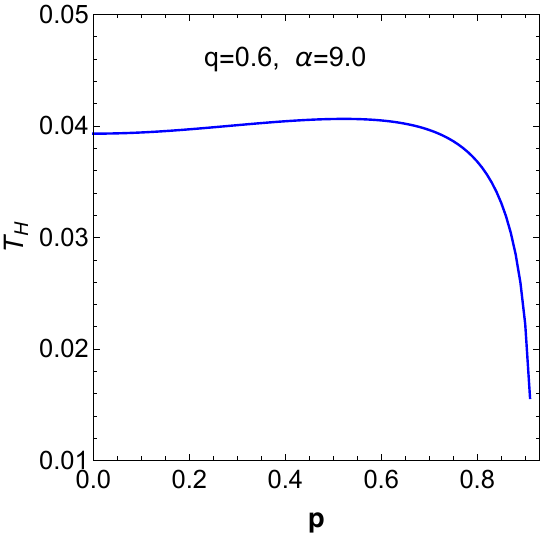}\;\;\;\;\includegraphics[width=5cm ]{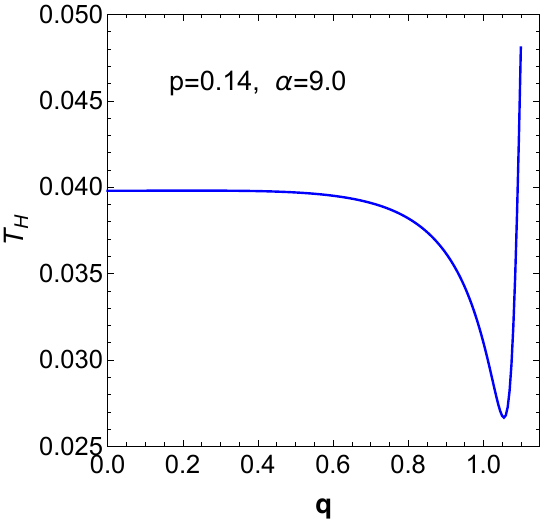}\;\;\;\;
\includegraphics[width=5.3cm ]{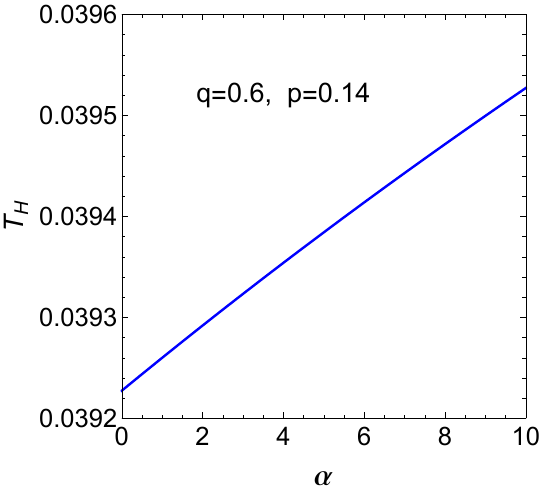}
\caption{Hawking temperature of the dyonic black hole (\ref{metric0}) with different parameters. Here we set $M=1$. }\label{hawte}
\end{figure}
In Fig.\ref{hawte}, we also present changes of Hawking temperature of the dyonic black hole (\ref{metric0}) with different parameters. It is well known that the black hole with higher Hawking temperature has higher evaporation rate and then has shorter lifetime. Fig.\ref{hawte} shows that the Hawking temperature  of the dyonic black hole (\ref{metric0}) increases with the parameter $\alpha$ and then its lifetime decreases for fixed $p=0.14$ and $q=0.6$. With the increase of $p$, we can obtain that the lifetime of the black hole first decreases and then increases for fixed  $q=0.6$ and $\alpha=9.0$.  With the increase of $q$, for $p=0.14$ and $\alpha=9.0$, the lifetime of the black hole first decreases and then increases, finally decreases again.

In order to study the equilibrium non-selfgravitating accretion tori, we must discuss the marginally stable orbit
 and the marginally bound orbit obtained from the metric (\ref{metric0}), which are two essential quantities for determining the thick disk model in a given spacetime. Lets us now focus on the equatorial geodesics with $\theta =\frac{\pi}{2}$ for a test timelike particle. Combining the two Killing vectors of the black hole background (\ref{metric0}), one can obtain two conserved quantities of particle moving along the geodesics, namely, its energy and angular momentum,
\begin{equation}
\mathcal{E}=-g_{tt} \dot{t}, \quad\quad \quad   L=g_{\phi \phi} \dot{\phi}.
\end{equation}
With these conserved quantities, the motion equation of the timelike particle moving in the equatorial plane can be simplified as
\begin{equation}
\dot{r}^2=\left(\mathcal{E}^2-V_{e f f}\right),
\end{equation}
where the effective potential is
\begin{equation}
V_{e f f}=\bigg(\frac{L^2}{r^2}+1\bigg)\bigg[1-\frac{2 M}{r}+\frac{p^2}{r^2}+\frac{q^2}{ r^2}{ }_2 F_1\left(\frac{1}{4}, 1 ; \frac{5}{4} ;-\frac{4  p^2 \alpha}{r^4}\right)\bigg].
\end{equation}
The circular orbit of a timelike particle is determined by $V_{ eff}=\mathcal{E}^2$ and $V'_{ eff}=0$. The stable circular orbits also  meet the second-order derivative of the effective potential $V_{e f f}^{\prime \prime}>0$.
The marginally bound orbit is the innermost unstable circular orbit for a timelike particle, which is determined by $V_{ eff}=1$ and $V'_{ eff}=0$.
\begin{figure}[ht]
\includegraphics[width=5.5cm ]{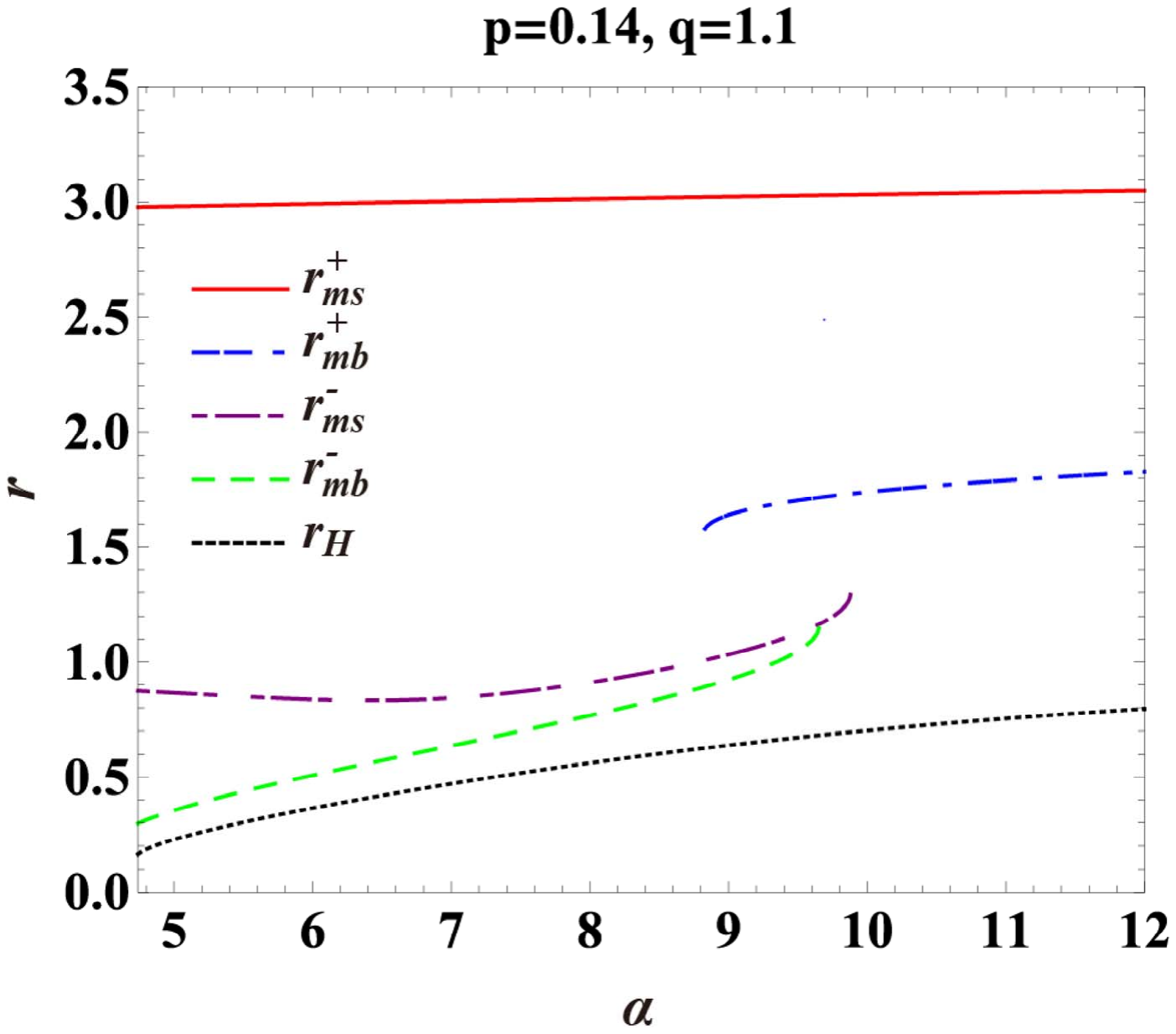}\includegraphics[width=5.5cm ]{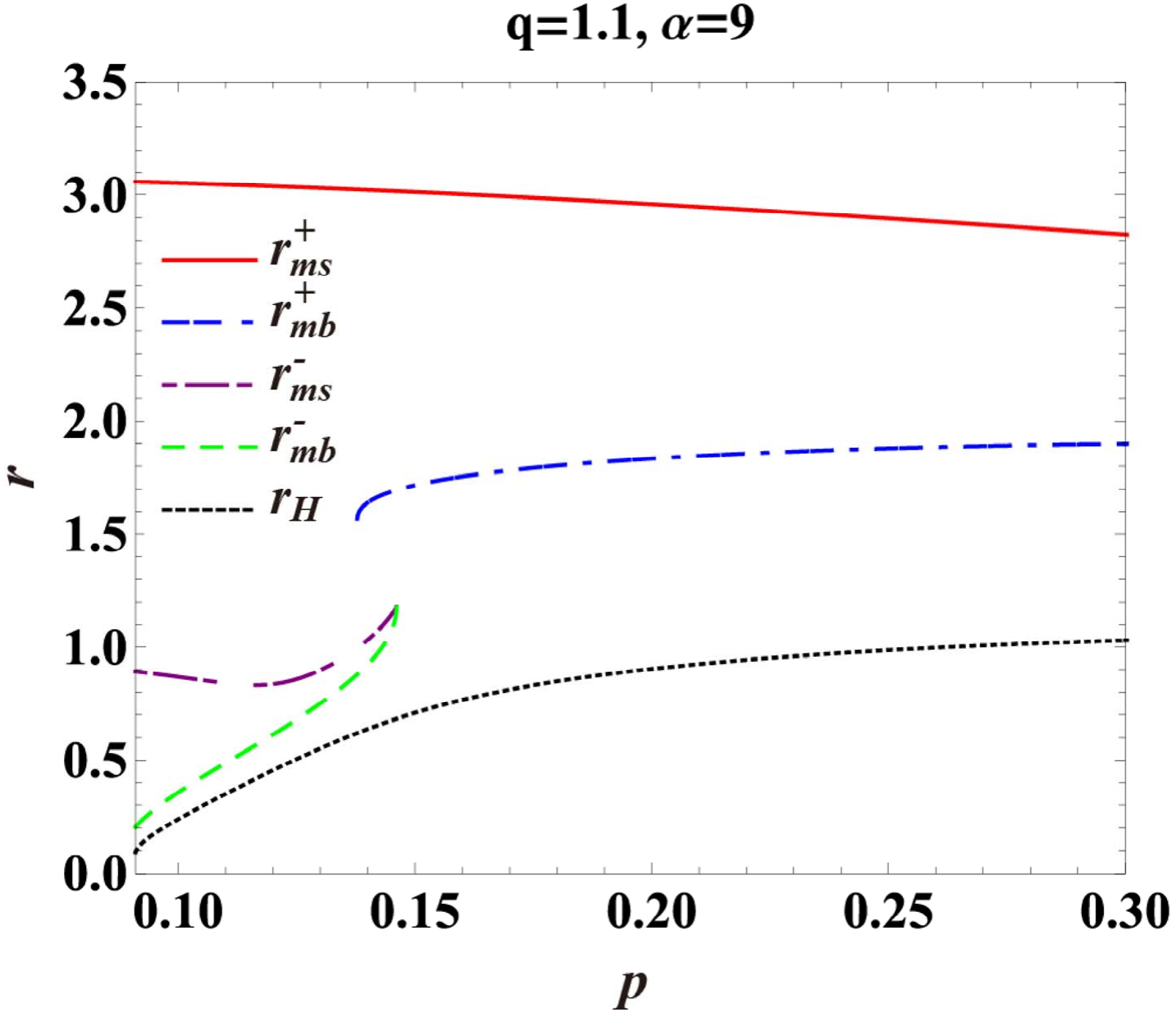}\includegraphics[width=5.5cm ]{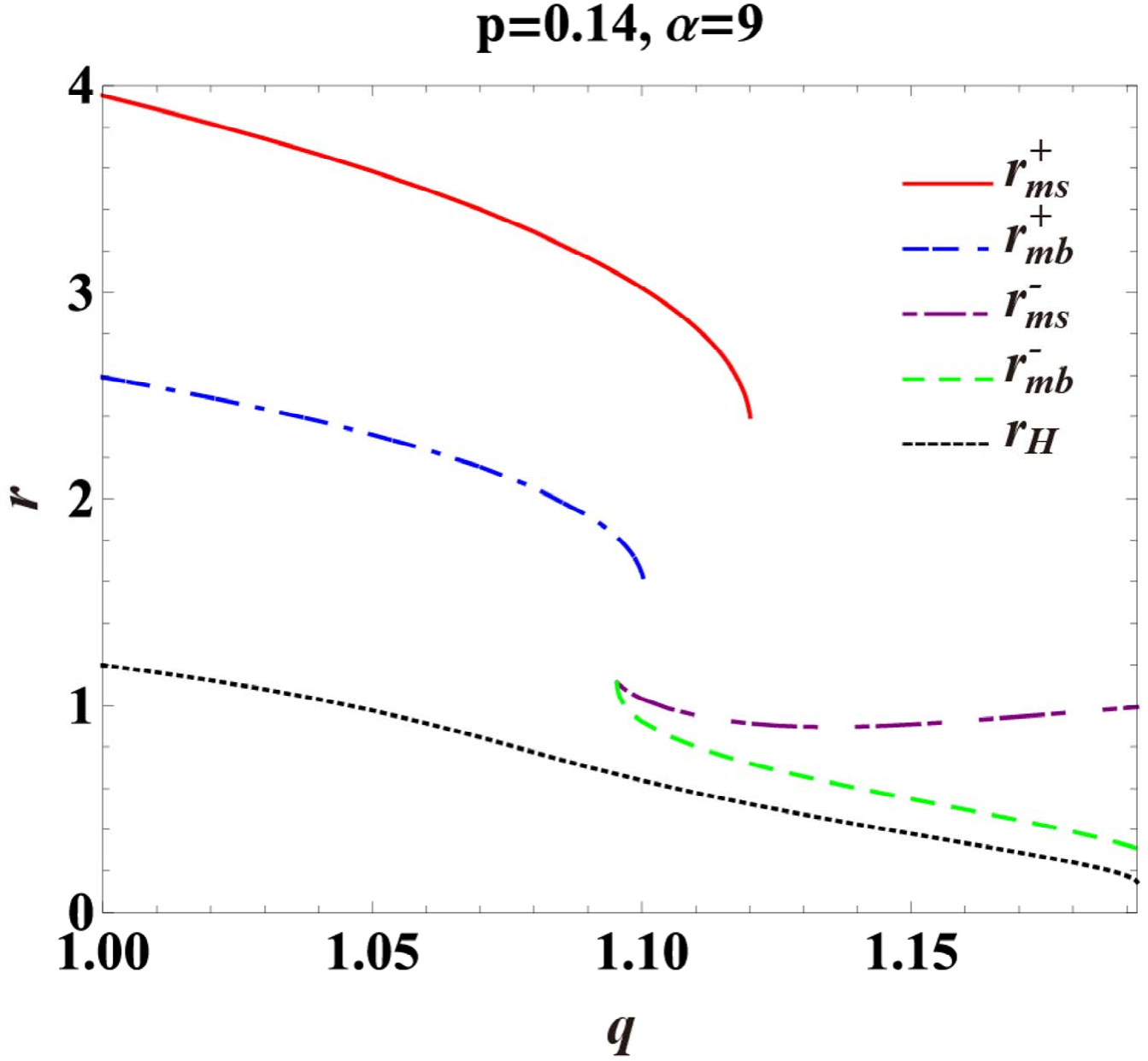}
\caption{ Effects of the parameters $\alpha$, $p$ and $q$ on the marginally stable orbit $r_{ms}$, the marginally bound orbit  $r_{mb}$ and the outer event horizon radius $r_H$ of the black hole. Here, we set $M=1$. }\label{ffms}
\end{figure}
Fig.(\ref{ffms}) shows effects of the parameters $\alpha$, $p$ and $q$ on the marginally stable orbit $r_{ms}$, the marginally bound orbit  $r_{mb}$ and the outer event horizon of the black hole. Theoretically, the choice of these parameters for a black hole is arbitrary as long as its event horizon exists. The main motivation of choosing such set of parameters is that there are multiple horizons for the black hole and the corresponding complex dynamical properties of test particles could yield richer configurations of thick equilibrium tori. Moreover, for fixed $p$, $q$ and $\alpha$, the relationship $2Q_eQ_m/\hbar=n$ ( where $n$ is an integer ) can also be ensured by adjusting the factor $\alpha_1$ or $\alpha_2$ because $Q_eQ_m=pq/\alpha_1
=pq\sqrt{\alpha}/\sqrt{\alpha_2}$.
From Fig.(\ref{ffms}), for fixed $p=0.14$ and $q=1.1$, we find that there exist two marginally stable orbits $r^{+}_{ms}$ and $r^{-}_{ms}$ in certain range of $\alpha$. With the increase of $\alpha$, the $r^{+}_{ms}$ monotonically increases, but $r^{-}_{ms}$ first decreases and then increases. With the further increase of $\alpha$, the inner marginally stable orbit $r^{-}_{ms}$ vanishes and the outer marginally stable orbit $r^{+}_{ms}$ still exists. For the marginally bound orbit, as $\alpha<8.830$ for the fixed $p=0.14$ and $q=1.1$, there exist only a marginally bound orbit and its radius $r^{-}_{mb}$ increases with $\alpha$. As $\alpha>8.830$, there exist another marginally bound orbit with larger radius $r^{+}_{mb}$. As $\alpha$ increases up to $9.683$, the inner marginally bound orbit disappears.  For fixed $q=1.1$ and $\alpha=9$, we find the changes of $r^{-}_{ms}$, $r^{\pm}_{mb}$ and $r_H$ with $p$ are similar to their changes with $\alpha$, but the $r^{+}_{ms}$ decreases with $p$. For fixed $p=0.14$ and $\alpha=9$, we find that $r^{+}_{ms}$, $r^{\pm}_{mb}$ and $r_H$ decrease with the electric charge parameter $q$, but $r^{-}_{ms}$ first decreases and then increases. These complex properties of particle's motion orbits are helpful to probe possible double equilibrium tori around the dyonic black hole. Moreover, we find that the orbit radii $r^{\pm}_{ms}$ and $r^{\pm}_{mb}$ exist simultaneously only within a narrow range of parameters $p$, $q$ and $\alpha$. In the Kerr black hole spacetime, it is found that there also exist the outer and inner marginally stable orbits $r^{+}_{ms}$, $r^{-}_{ms}$,  the outer and inner marginally bound orbits $r^{+}_{mb}$, $r^{-}_{mb}$. This is caused by the black hole spin and the test particle co-rotating or  counter-rotating  around the black hole. These co-rotating and counter-rotating rings yield some equilibrium configurations in ringed
accretion disks \cite{2015aj229240P,2017aj229225P}. It is differen from those in the static dyonic black hole spacetime (\ref{metric0}) where the existences of these characterized orbits and the corresponding double equilibrium tori are caused by gravitational effects from quasi-topological electromagnetism rather than the black hole rotation. Moreover, in the Kerr black hole, the orbit radii $r^{\pm}_{ms}$ and $r^{\pm}_{mb}$ always exist simultaneously for arbitrary non-zero spin value. This is another difference from that in the case of the dyonic black hole with quasi-topological electromagnetism (\ref{metric0}).

Let us now to briefly review the Polish doughnuts model about the equilibrium torus of  electrically neutral and stationary rotating fluids around a black hole
 \cite{1974AcA2445A,1971AcA2181A,1978A&A63209K,1978AA63221A,1980AcA301J,1980A&A8823P,1980ApJ242772A,
1982MitAG5727P,1982ApJ253897P}. For the perfect fluid, its stress energy tensor can be expressed as
\begin{equation}
T^{\mu \nu}= w u^\mu u^\nu + P g^{\mu \nu}\label{eq8}
\end{equation}
where $P$ and $w$ are the pressure and the enthalpy density of the fluid, respectively. $u^\mu$ is the four velocity of the fluid particle, which satisfies the relationship $u_\mu u^\mu=-1$.  In the Polish doughnuts model, the fluid in the thick disk is assumed to be a barotropic perfect fluid with positive pressure and its self-gravity is
negligible so that the influence of  disk on background spacetime is negligible. Moreover, the fluid is assumed to be axisymmetric and stationary, and the rotation of perfect fluid is restricted to be in the
azimuthal direction. With these assumptions, the four velocity and the stress energy tensor of the perfect fluid can be expressed as
\begin{eqnarray}
u^{\mu}&=&(u^t,0,0,u^{\phi}).
\end{eqnarray}
From the conservation for the perfect fluid $\nabla_\mu T_\nu^\mu=0$,  one can obtain \cite{1974AcA2445A,1971AcA2181A,1978A&A63209K,1978AA63221A,1980AcA301J,1980A&A8823P,1980ApJ242772A,
1982MitAG5727P,1982ApJ253897P}
\begin{equation}
-\nabla_{\nu} \ln u_t+\frac{\Omega \nabla_{\nu} l}{1-l \Omega}=\frac{1}{w} \nabla_{\nu} P. \label{eq10}
\end{equation}
In the background spacetime (\ref{metric0}), the redshift factor of the fluid particle $u_t$  has a form
\begin{equation}
u_t=\sqrt{\frac{g_{t t} g_{\phi \phi}}{l^2 g_{t t}+g_{\phi \phi}}}.\label{eq9}
\end{equation}
 The quantities $\Omega$ and $l$ are respectively the angular velocity and specific angular momentum of the fluid particle
\begin{equation}
 \Omega=\frac{d \phi}{d t}, \quad \quad  \quad l=\frac{L}{\mathcal{E}}=-\frac{g_{\phi\phi}\dot{\phi}}{g_{tt}\dot{t}}=-\frac{g_{\phi\phi}\Omega}{g_{tt}}.\label{Omegal}
\end{equation}
For a barotropic fluid,  the enthalpy is a function of $p$ and the right side of Eq.(\ref{eq10})
is an exact differential. According to the von Zeipel theorem \cite{1974AcA2445A,1971AcA2181A,1978A&A63209K,1978AA63221A,1980AcA301J,1980A&A8823P,1980ApJ242772A,
1982MitAG5727P,1982ApJ253897P}, one can get $\Omega=\Omega(l)$ and then
can obtain a solution by integrating Eq.(\ref{eq10}) to obtain
\begin{equation}
W_{eff}-W_{in}=\ln \left|u_t\right|-\ln \left|\left(u_t\right)_{\mathrm{in}}\right|-\int_{l_{\mathrm{in}}}^{l}\left(\frac{\Omega}{1-\Omega l'}\right) d l',
\end{equation}
The subscript ``in" denotes that the quantity is evaluated at
at the inner edge of the disk. The potential $W_{eff}$ is given by
\begin{equation}
W_{eff}-W_{in}=\int_0^{P}\frac{dP}{w},\label{wwdp}
\end{equation}
which determines the topologies of
equipotential surfaces in the disk. As in \cite{Gimeno-Soler:2017qmt}, we assume that a barotropic equation of state takes a form $P=K w^{\kappa}$, where $K$ and $\kappa$ are constants. Inserting into Eq. (\ref{wwdp}), one can easily obtain
\begin{equation}
W_{eff}-W_{in} + \frac{\kappa}{\kappa-1}\frac{P}{w}=0.\label{wwinp}
\end{equation}
Thus, once the potenial $W_{eff}$ and the specific enthalpy $h=w/\rho$ are given, one can get the enthalpy density $w$ and the rest-mass density $\rho$ distribution in the disk. Here, we set  $h=1$ for simplicity \cite{Gimeno-Soler:2017qmt}. Moreover, we here focus on only the case in which the fluid
has a constant specific angular momentum $l$ because $dl=0$ and the fluid
angular velocity $\Omega$ in Eq.(\ref{Omegal} ) becomes a function only related to spacetime metric,  which makes the
calculation of equilibrium tori particularly simple.  In this simple model, the potential $W_{eff}$ can be further simplified as
$W_{eff}=\ln|u_t|$. Such stationarily rotating perfect-fluid tori with this property are known as
``Polish doughnuts".

\section{EQUILIBRIUM TORI around  the dyonic black hole}

\begin{figure}[ht]
\includegraphics[width=5.5cm ]{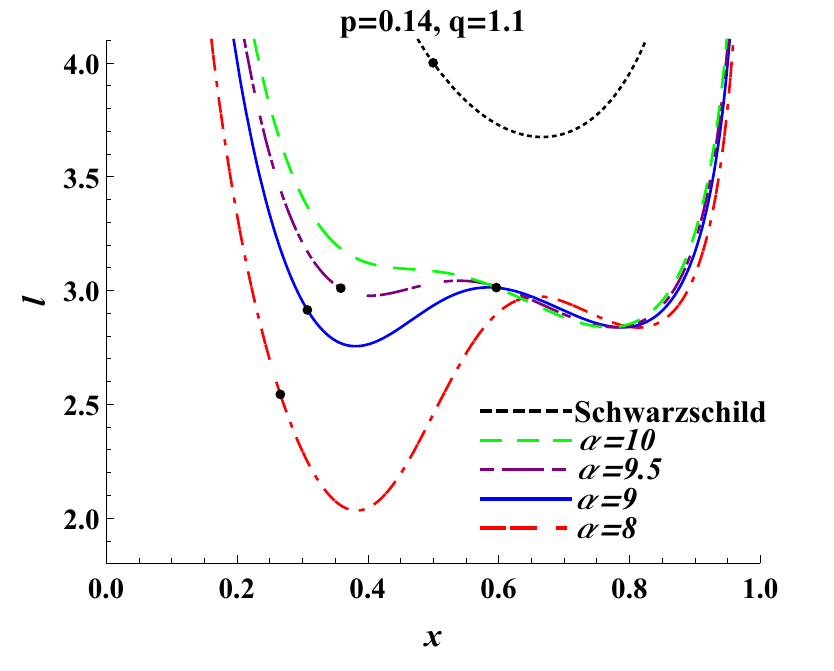}
\includegraphics[width=5.5cm ]{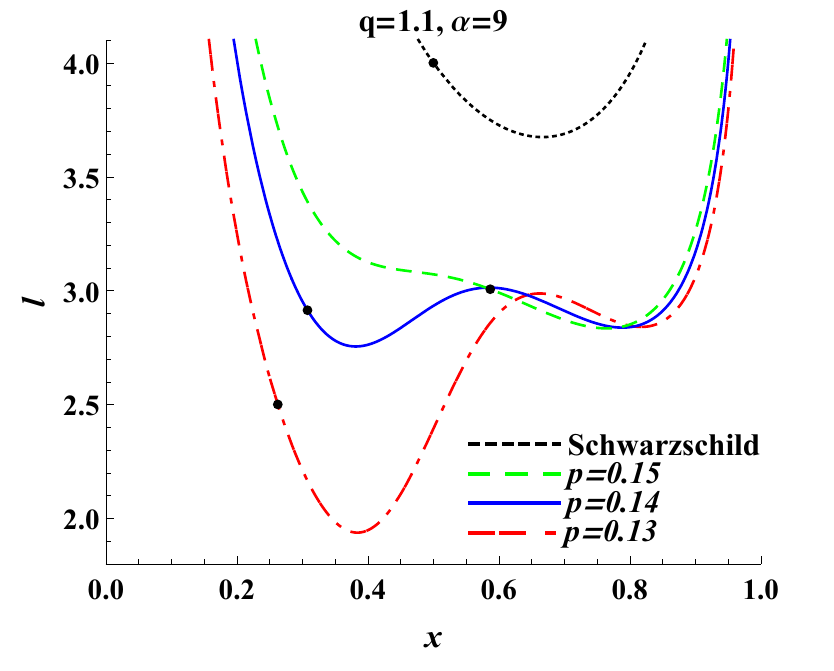}
\includegraphics[width=5.5cm ]{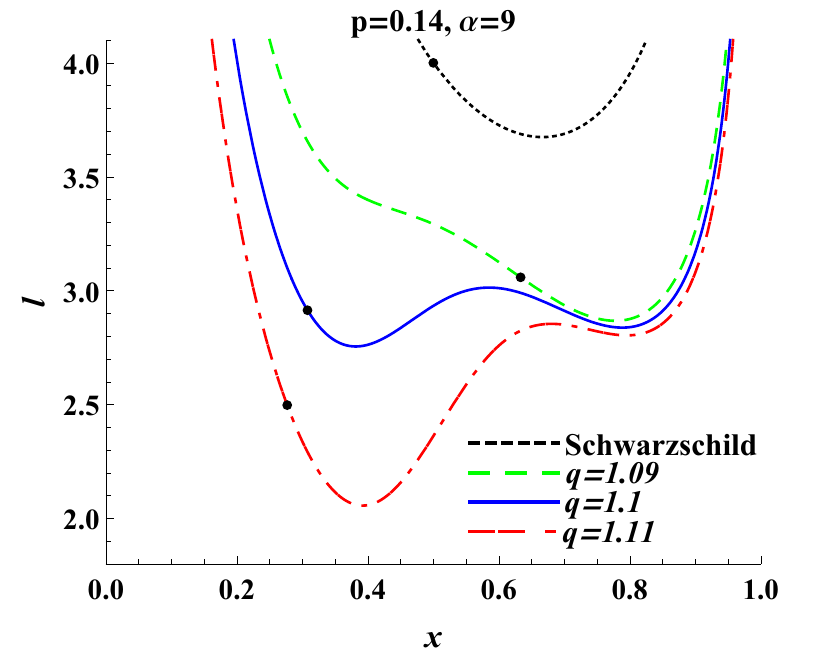}
\caption{ Changes of the specific angular momentum $l$ with the circular orbit radius $x=1-\frac{r_{H}}{r}$ for different parameters, where $r_H$ is the outermost horizon of the black hole. The black dots indicate the values of $l_{mb}$.}\label{f1}
\end{figure}

\begin{figure}[ht]
\includegraphics[width=5cm ]{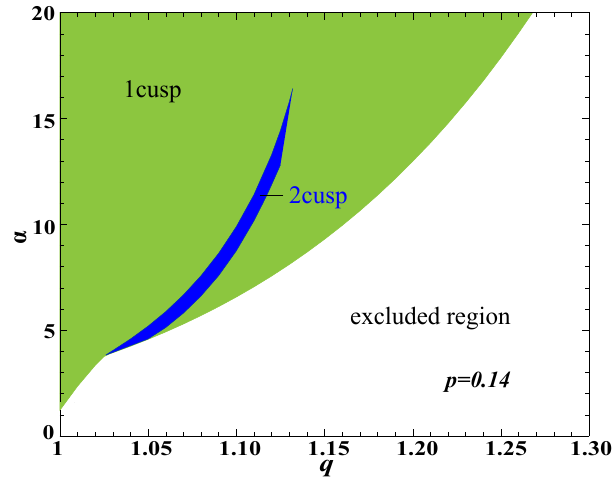}\;
\includegraphics[width=5cm ]{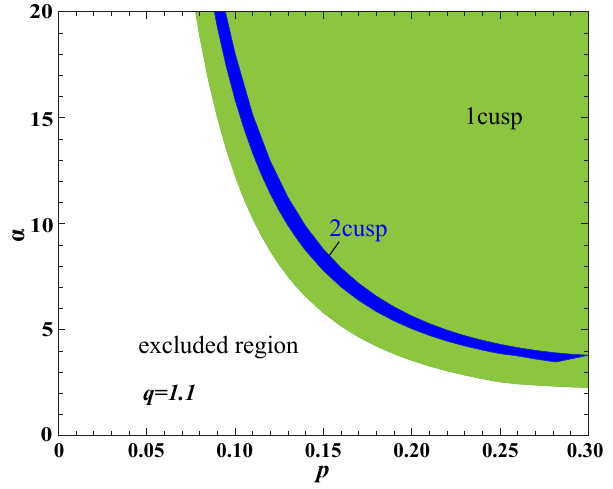}\;
\includegraphics[width=5cm ]{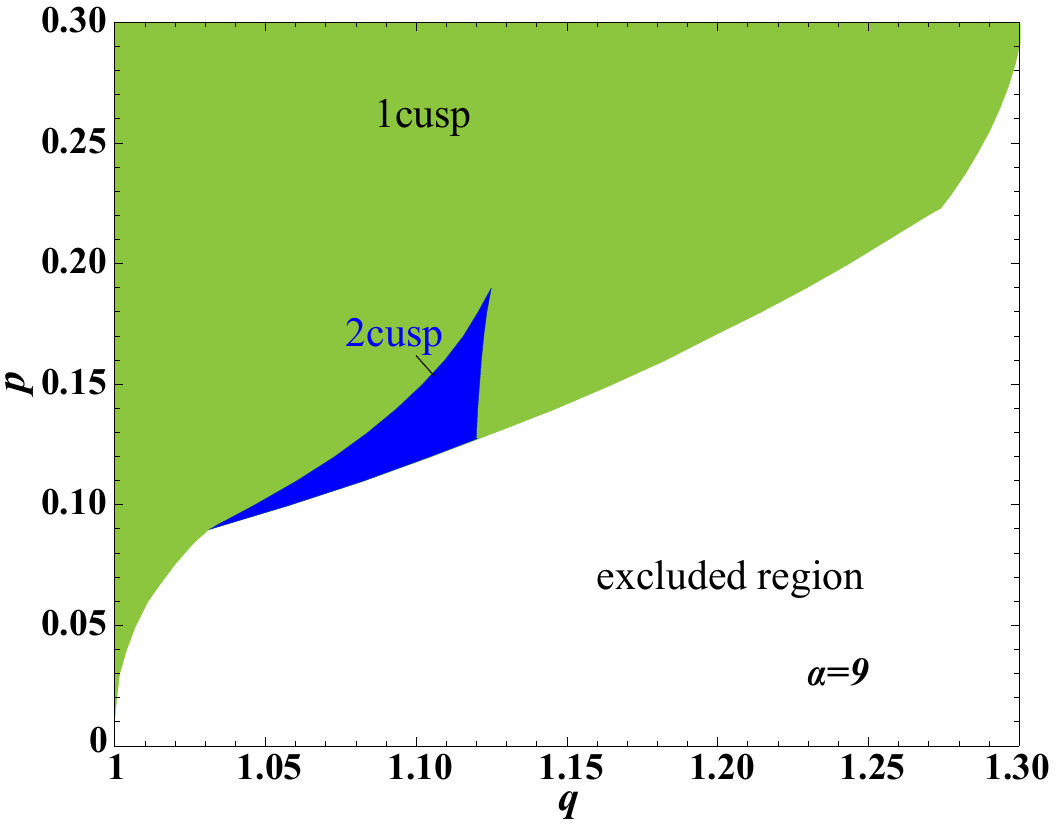}\;
\caption{Multidimensional parameter space for tori solutions around the dyonic black hole (\ref{metric0}).
In the blue region there may exist double tori and two cusps. The green region indicates that there is only a single torus  and one cusp, and the white region indicates that there is no equilibrium torus.}\label{f2}
\end{figure}

In this section, we will study the electrically neutral equilibrium tori around the dyonic black hole and probe effects of the electric, magnetic charge and coupling parameters on the equilibrium tori.
As in the previous discussion, the specific angular momentum $l$ plays an important role in the potential $W_{eff}$ and  has a great impact on the fluid equilibrium tori around black holes \cite{1974AcA2445A,1971AcA2181A,1978A&A63209K,1978AA63221A,1980AcA301J,1980A&A8823P,1980ApJ242772A,
1982MitAG5727P,1982ApJ253897P}. Figure.(\ref{f1}) shows the changes of the specific angular momentum $l$ of the fluid particles with the rescaled electric, magnetic and coupling parameters $(q,\;p,\;\alpha)$ in the dyonic black hole spacetime (\ref{metric0}).
When the specific angular momentum is equal to that at the marginally bound
orbit, $l=l_{mb}$,  one can find that a cusp is located at the marginally closed surface that
just extents to infinity. As $l>l_{mb}$,  the outermost equipotential surfaces are connect with the event
horizon  and the tori are effectively accreting without crossing the cusp. Actually, the maximum of
the potential $W_{eff}$ in this case is larger than the corresponding value at
spatial infinity and there is no stable torus in these regions. As the specific angular momentum is less than
the value of the marginally stable orbit, $l<l_{ms}$, there is no equilibrium torus. From Fig.\ref{f1}, one can find that the curves $l(r)$ for different parameters can be classified as three types: (i) The curve $l(r)$ has only a minimum value $l_{ms}$ and the specific angular momentum at
the marginally bound
orbit $l_{mb}$ is larger than the minimum value, $l_{mb}>l_{ms}$, and then there may be only a single tori around the black hole, which is similar to that in the Schwarzschild black hole spacetime. (ii) The curve $l(r)$ has two minimum values, $l^{+}_{ms}$ and $l^{-}_{ms}$, and $l_{mb}$ is larger than both of the two minimum values, $l_{mb}>l^{\pm}_{ms}$, then there may be double tori around the black hole,  which does not appear in the Schwarzschild spacetime. (iii)  The curve $l(r)$ has two minimum values, $l_{mb}$ is in between these two minimum values, i.e., $l^{+}_{ms}>l_{mb}>l^{-}_{ms}$,  then there may still be only a torus in this case.
In Fig.\ref{f2}, we show the parameter space for tori solutions around the dyonic black hole (\ref{metric0}). The green-shaded region corresponds to the case where the single-torus solutions can be found, the blue-shaded region is the region in which double-tori solutions are possible.

\begin{figure}[ht]
\includegraphics[width=5cm ]{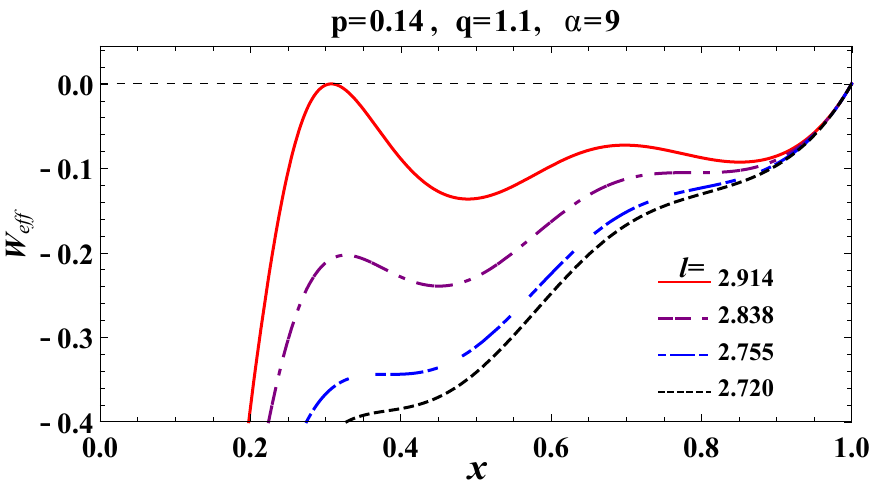}\includegraphics[width=5cm ]{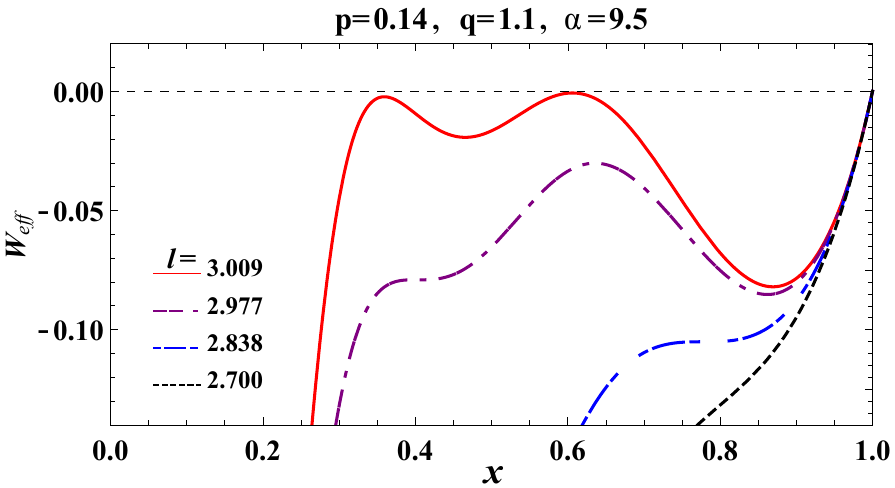}\includegraphics[width=5cm ]{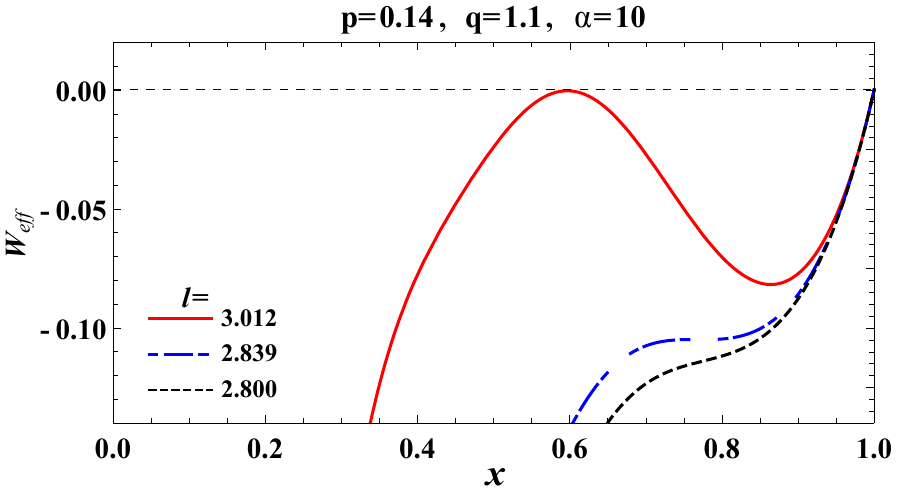}
\caption{Changes of the effective potential $W_{eff}$ in the equatorial plane with the radial coordinate $x$ for different specific angular momentum $l$ in the dyonic black hole  spacetime (\ref{metric0}).}\label{f3}
\end{figure}

Fig.\ref{f3} presents the changes of $W_{eff}$ in the equatorial plane with the radial coordinate $x=1-r_H/r$ for different specific angular momentum $l$. The cusps and the tori centres are respectively located at the positions of the local minima and maxima of $W_{eff}$. From Fig.\ref{f3}, one can find that the cusps move toward the black hole with the increase of $l$, while the tori centres move outward. In the dyonic black hole spacetime (\ref{metric0}) with $p=0.14$, $q=1.1$ and $\alpha=10$, one can find that the potential $W_{eff}$ has only one local maximum and  minimum for certain specific angular momentum $l$.
This means that there exist only a cusp and single equilibrium torus for the fluid with the constant specific angular momentum $l$ around the black hole. In the cases with $p=0.14$, $q=1.1$ and $\alpha=9$ or $\alpha=9.5$, we find that there exist two local maxima  at $r=r^{\pm}_{cusp}$ and two minima at $r=r^{\pm}_{max}$ for the potential $W_{eff}$ in the some range of $l$. Thus, there are two cusps and two equilibrium torus around the black hole. Moreover, from Fig.\ref{f3}, we also note  that for the dyonic black hole with parameters $p=0.14$, $q=1.1$ and $\alpha=9$, there are double equilibrium tori if $2.838<l<2.914$, and the centres of the inner and outer tori are located at $r=r^{-}_{max}$ and $r=r^{+}_{max}$. However, as $l$ decreases down to in the range $2.755<l<2.838$, the outer torus with the centre $r=r^{+}_{max}$ vanishes, but the inner torus is remained, and then there is a single equilibrium torus around the black hole. For the dyonic black hole with parameters $p=0.14$, $q=1.1$ and $\alpha=9.5$, we find that there also exist double equilibrium tori if $2.977<l<3.009$. Similarly, as $l$ decreases down to in the range
 $2.838<l<2.977$, there is a single equilibrium torus around the black hole. The difference from the former is that here the inner torus vanishes and the outer torus is remained.
\begin{figure}[ht]
\includegraphics[width=5cm ]{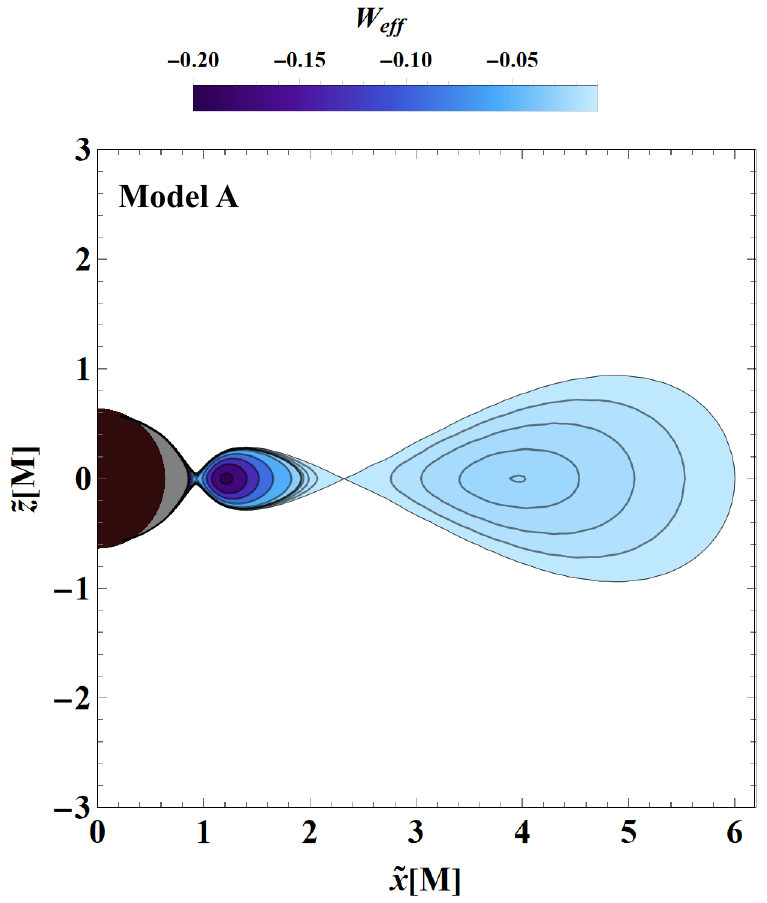}\;\includegraphics[width=5cm ]{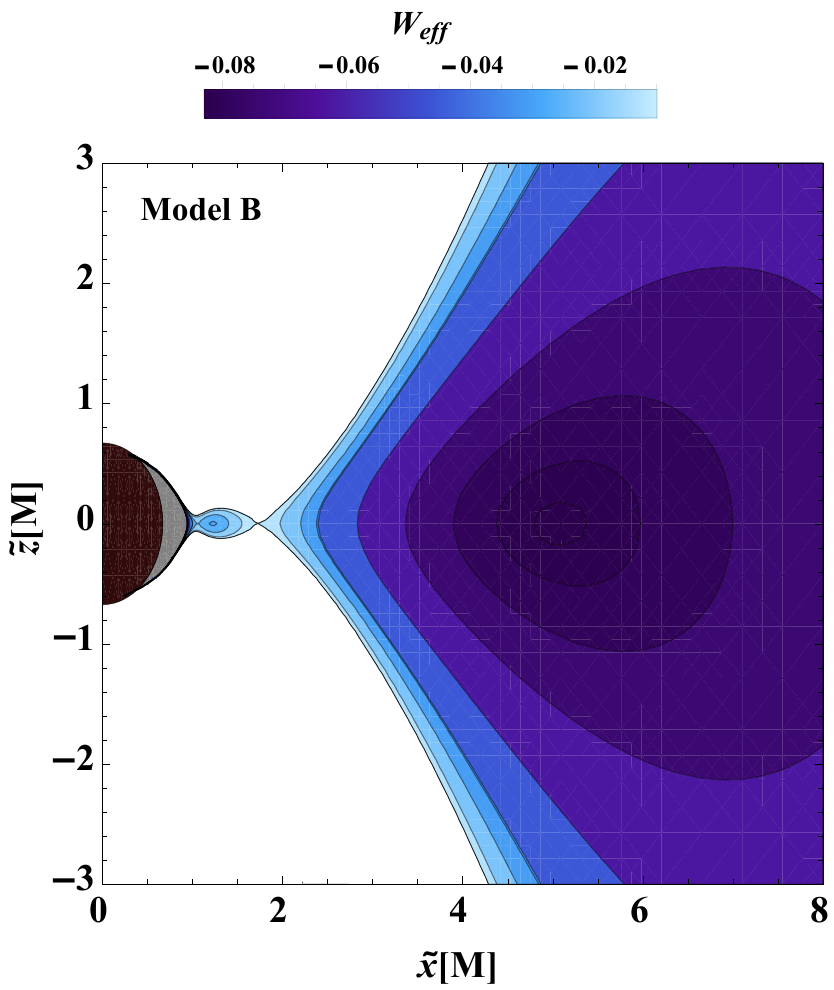}\;\includegraphics[width=5cm ]{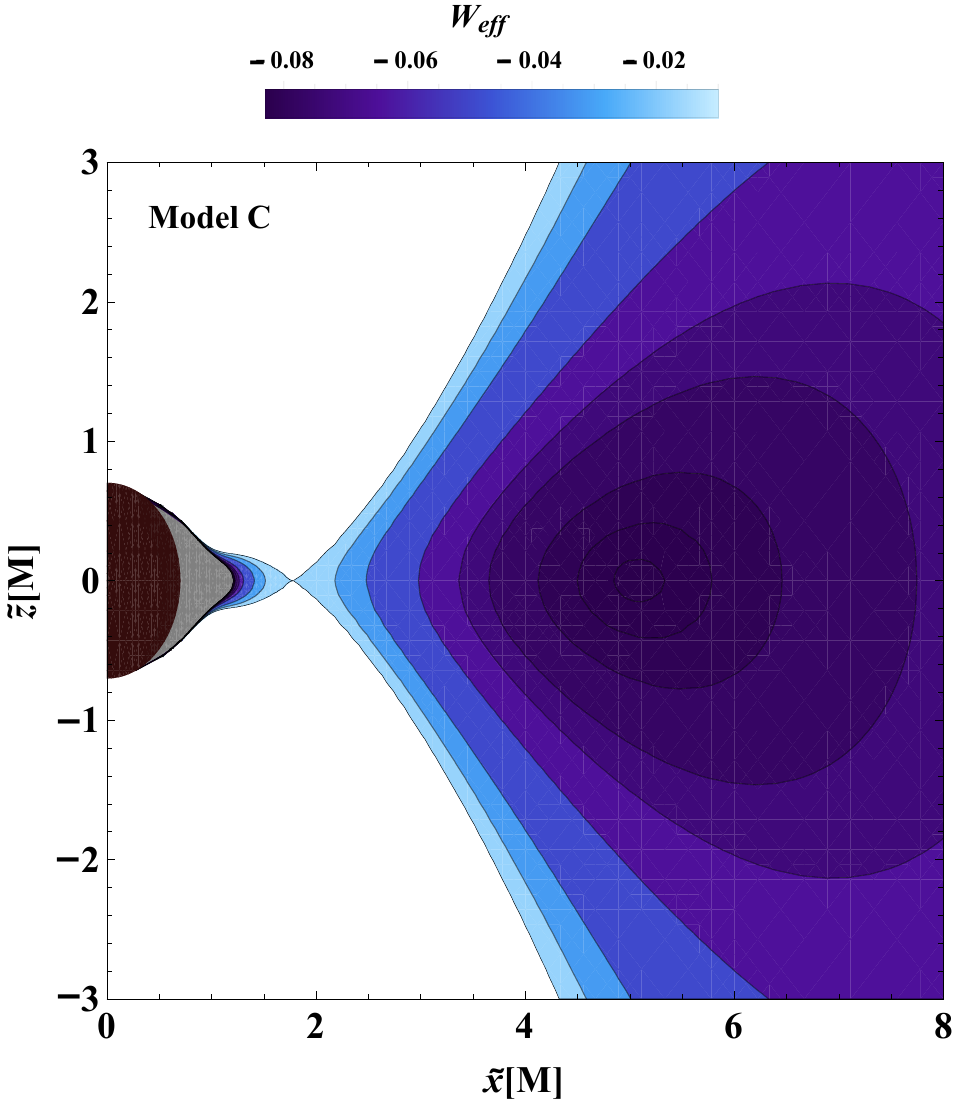}
\caption{Equipotential surfaces of the effective potential $W_{eff}$ shown with Cartesian coordinates $\tilde{x}=r\sin\theta\cos\phi$ and $\tilde{z}=r\cos\theta$ for black hole parameters $p=0.14$ and $q=1.1$. The specific angular momentum is set to $l=2.88$ in the left panel and is set to $l=3$ in the middle and right panels.  The parameter $\alpha$ is respectively set to $\alpha=9$, $9.5$ and $10$ from left to right.} \label{f5}
\end{figure}
\begin{figure}[ht]
\includegraphics[width=5cm ]{11zx9t.pdf}\;\includegraphics[width=5cm ]{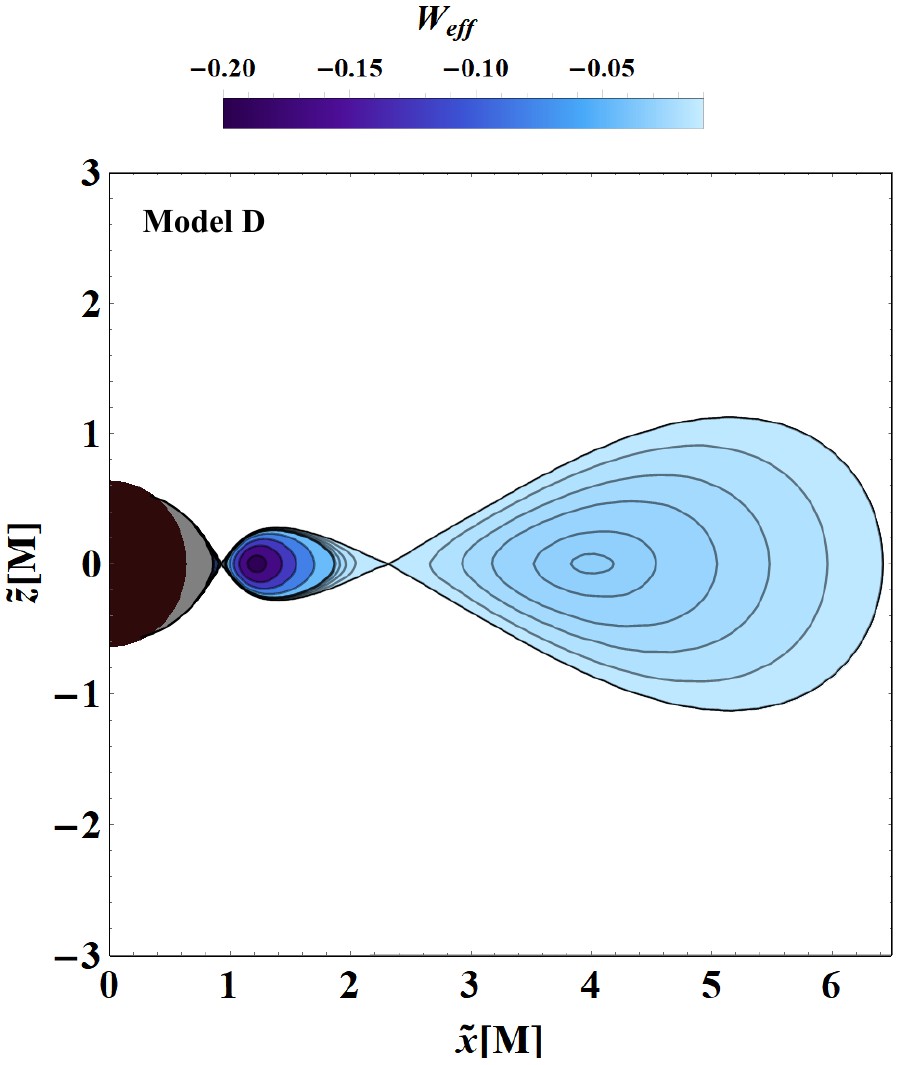}\;\includegraphics[width=5cm ]{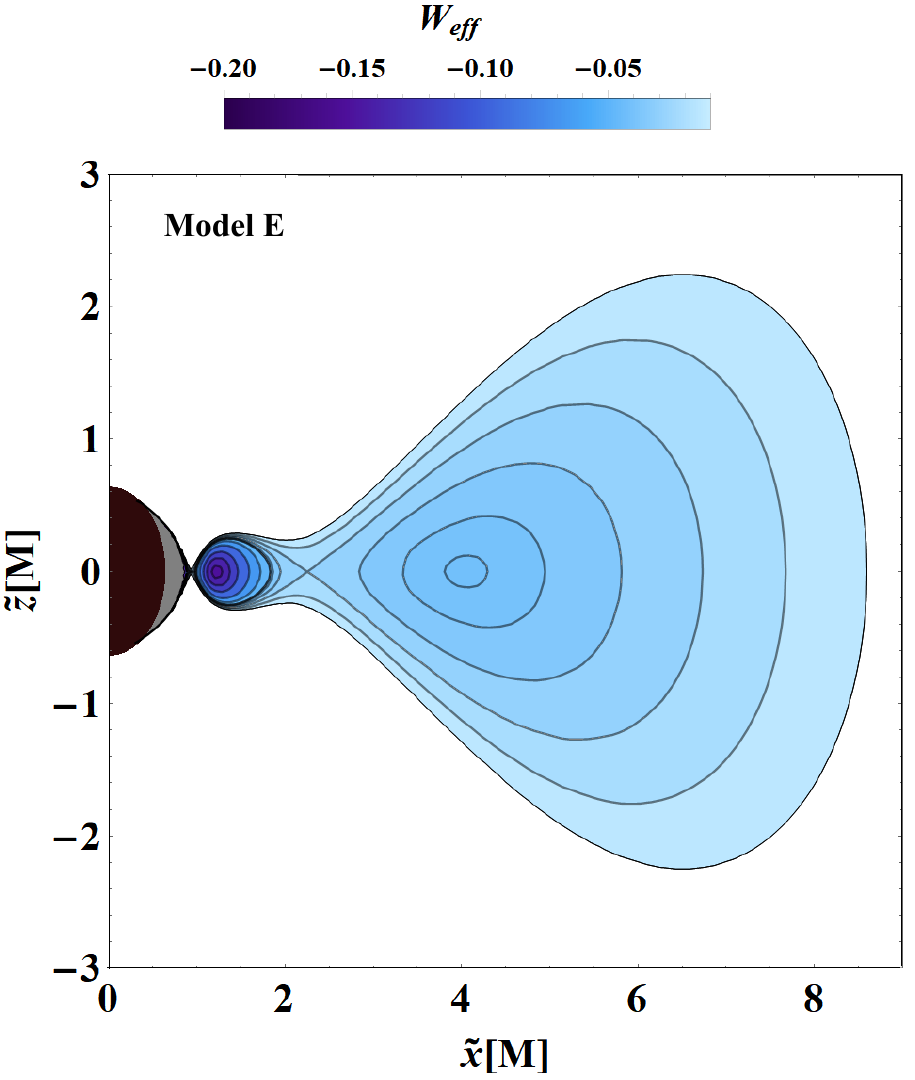}
\caption{Equipotential surfaces of the effective potential $W_{eff}$ shown with Cartesian coordinates $\tilde{x}=r\sin\theta\cos\phi$ and $\tilde{z}=r\cos\theta$ for black hole parameters $p=0.14$, $q=1.1$ and $\alpha=9$. The specific angular momentum are set to $l=2.88$, $l=2.885865$ and $l=2.89$ in the panels  from the left to the right, which respectively correspond to the Models A, D and E.} \label{f51}
\end{figure}
These are also shown in Figs. \ref{f5} and \ref{f51}, in which
the equipotential surfaces of the effective potential $W_{eff}$ are shown in Cartesian coordinates $\tilde{x}[M]$ and $\tilde{z}[M]$. The parameters of the models A-E for equilibrium tori are listed in Table \ref{ttable1}.
For the cases with double equilibrium tori, if the value of the potential $W_{eff}$ at the inner cusp is less than one at the outer cusp, i.e., $W_{eff}(r^{-}_{cusp})<W_{eff}(r^{+}_{cusp})$, we find that the equipotential surface of the inner cusp is inside the equipotential surface of the outer one. As shown in the left panel (Model A) in Fig.\ref{f51}, for the fluid particle with $W_{eff}(r^{-}_{cusp})<W_{eff}<W_{eff}(r^{+}_{cusp})$,  it can be accreted into the black hole if the particle is located inside the inner torus even if its equipotential is less than one at the outer cusp. However, if this fluid particle is located inside the outer torus, it can still keep equilibrium  although its potential is more than one at the inner cusp. These are different from that in the single cusp case in which the fluid particles can be accreted into black hole only if their  potentials $W_{eff}$ are larger than that of the cusp. As $W_{eff}(r^{-}_{cusp})=W_{eff}(r^{+}_{cusp})$, the inner and the outer cusps belong to the same equipotential surface, and the accretion occurs if the potential of the fluid particle $W> W_{eff}(r^{-}_{cusp})=W_{eff}(r^{+}_{cusp})$, which is shown in the middle panel (Model D) in Fig.\ref{f51}.
If $W_{eff}(r^{-}_{cusp})>W_{eff}(r^{+}_{cusp})$, we find that the fluid particles with $W_{eff}(r^{+}_{cusp})<W_{eff}<W_{eff}(r^{-}_{cusp})$ can move along the circular orbit and cannot be accreted into the black hole as shown in the right panel (Model E) in Fig.\ref{f51}, which is also different from that in the case with single cusp. In Figs. \ref{f5} and \ref{f51}, we also note that the fluid particles in the gray regions can not form equilibrium tori.
 Thus, in the double equilibrium tori case,  the occurrence of fluid particle accretion depends on the potential values at the two cusps and the position of fluid particle.
\begin{table}
\begin{tabular}{|c|c|c|c|c|c|c|c|c|c|c|c|}
\hline\hline
 \text { Model } & \quad $\alpha$  \quad  & \quad  $l$ \quad  & \quad $r_H$  \quad & \quad $r^+_{ cusp }$ \quad &\quad $r^-_{ cusp }$ \quad& \quad $r_{max}$ \quad &
 \quad $W_{ r^+_{cusp }}$  \quad \quad &
 \quad $W_{ r^-_{cusp }}$  \quad &  \quad $W_{r_{max}}$ \quad  & \quad $W_{in}$ \quad
 \\
\hline \text { A } & 9.0 & 2.88 & 0.639  & 2.328 &0.9332& 3.9206 & -0.08968 & -0.10311 & -0.09751  &  -0.08968 \\
\hline \text { B } & 9.5 & 3.00 & 0.672  & 1.747 &1.0589& 5.0720 & -0.00968 & -0.02596 & -0.08295  & -0.00968 \\
\hline \text { C } & 10.0 & 3.00 &  0.702 & 1.786 &-- & 5.0720  & -0.01182 &-- &-0.08295 & -0.01182 \\
\hline \text { D } & 9.0 & 2.885865& 0.639 & 2.286 &0.9313 & 3.9907  & -0.08702 & -0.08702 &-0.09662 & -0.08702 \\
\hline \text {E } & 9.0 & 2.89  &  0.639 & 2.258 & 0.9299 &4.0384  & -0.08507 & -0.07530 &-0.09600 & -0.07530 \\
\hline\hline
\end{tabular}
\caption{Parameters of the models A-E for equilibrium tori.}\label{ttable1}
\end{table}
\begin{figure}[ht]
\includegraphics[width=5.2cm ]{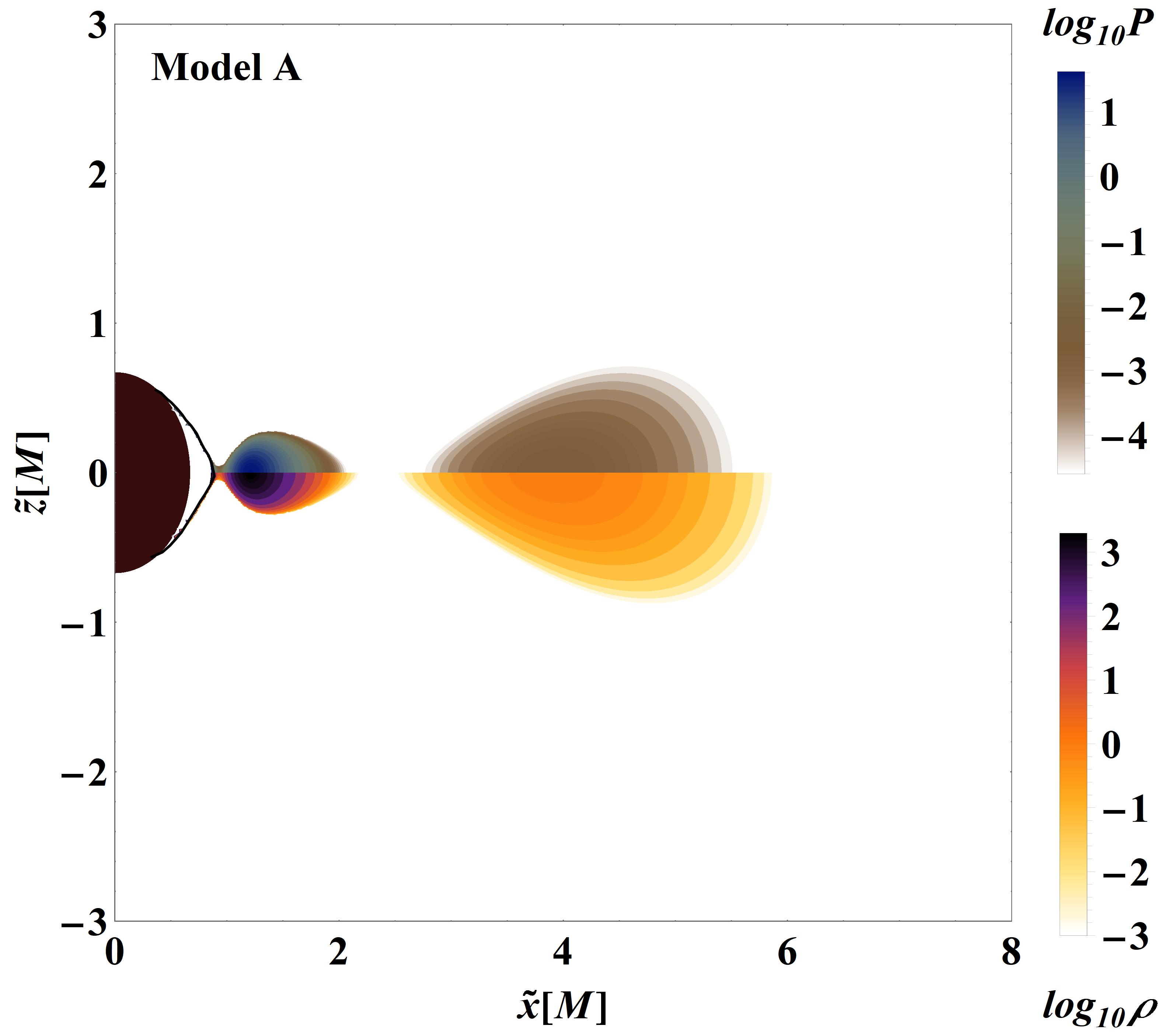}\;\includegraphics[width=5.2cm ]{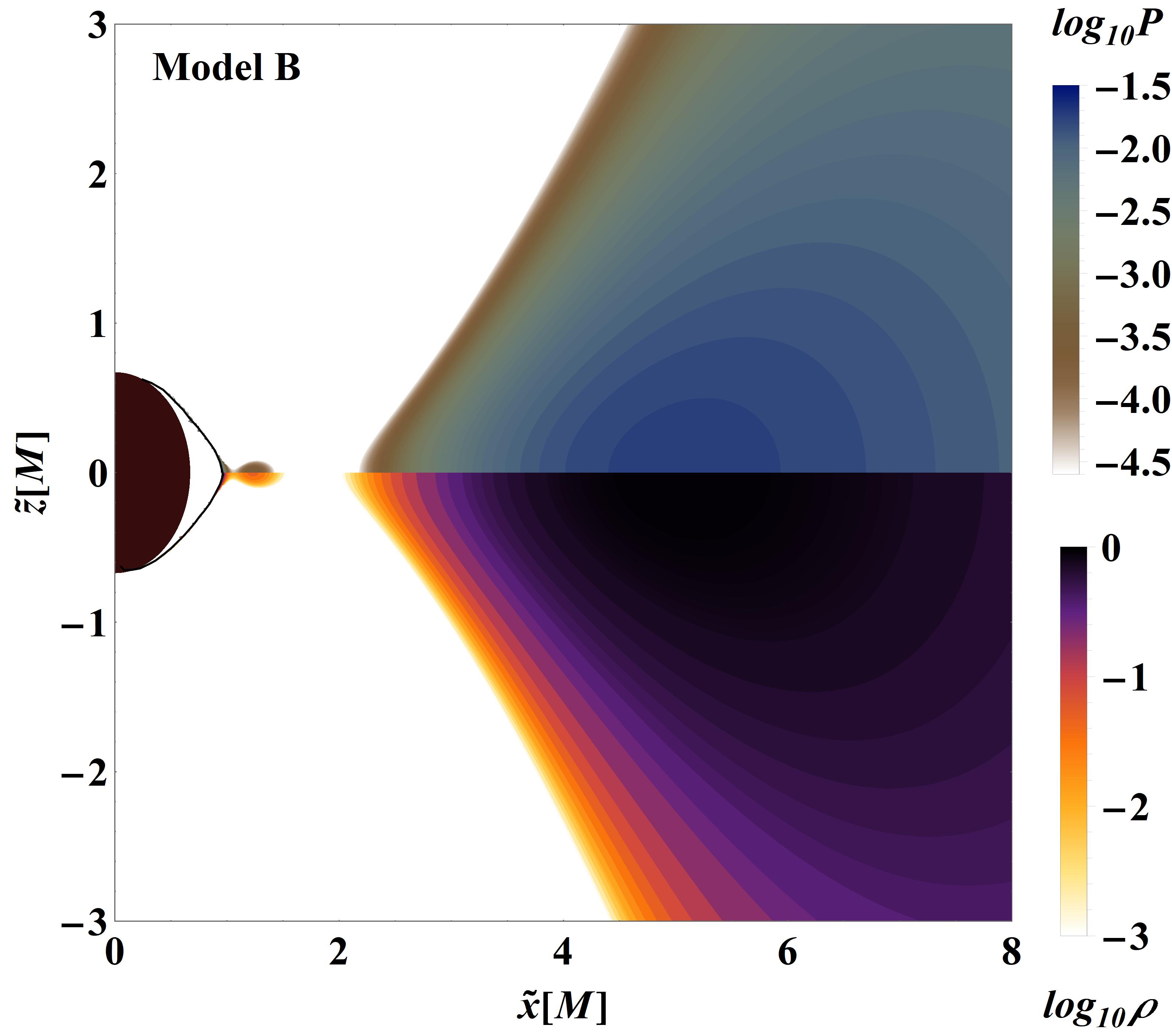}\;\includegraphics[width=5.2cm ]{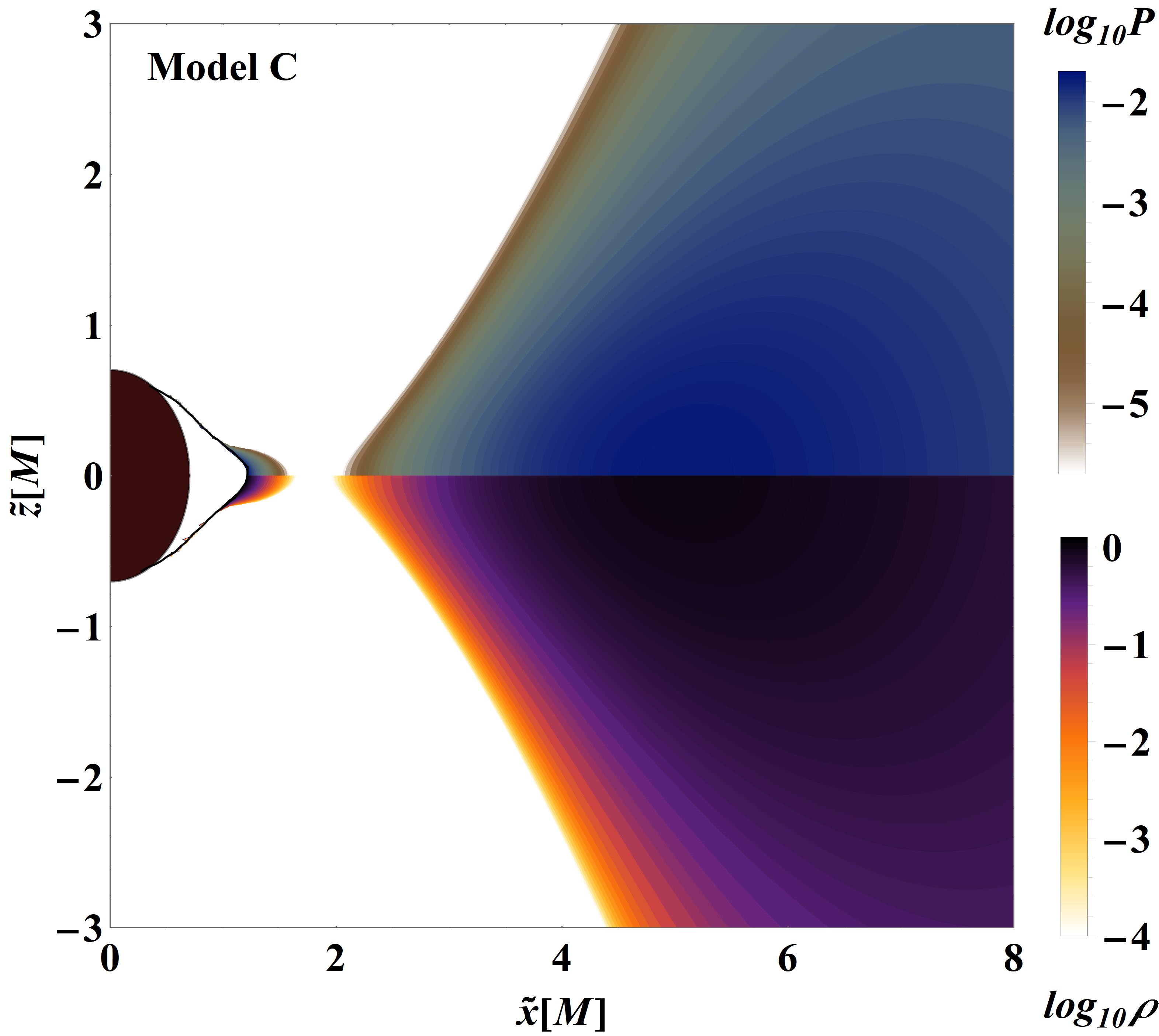}
\includegraphics[width=5.2cm ]{161a9288J.jpg}\;\includegraphics[width=5.2cm ]{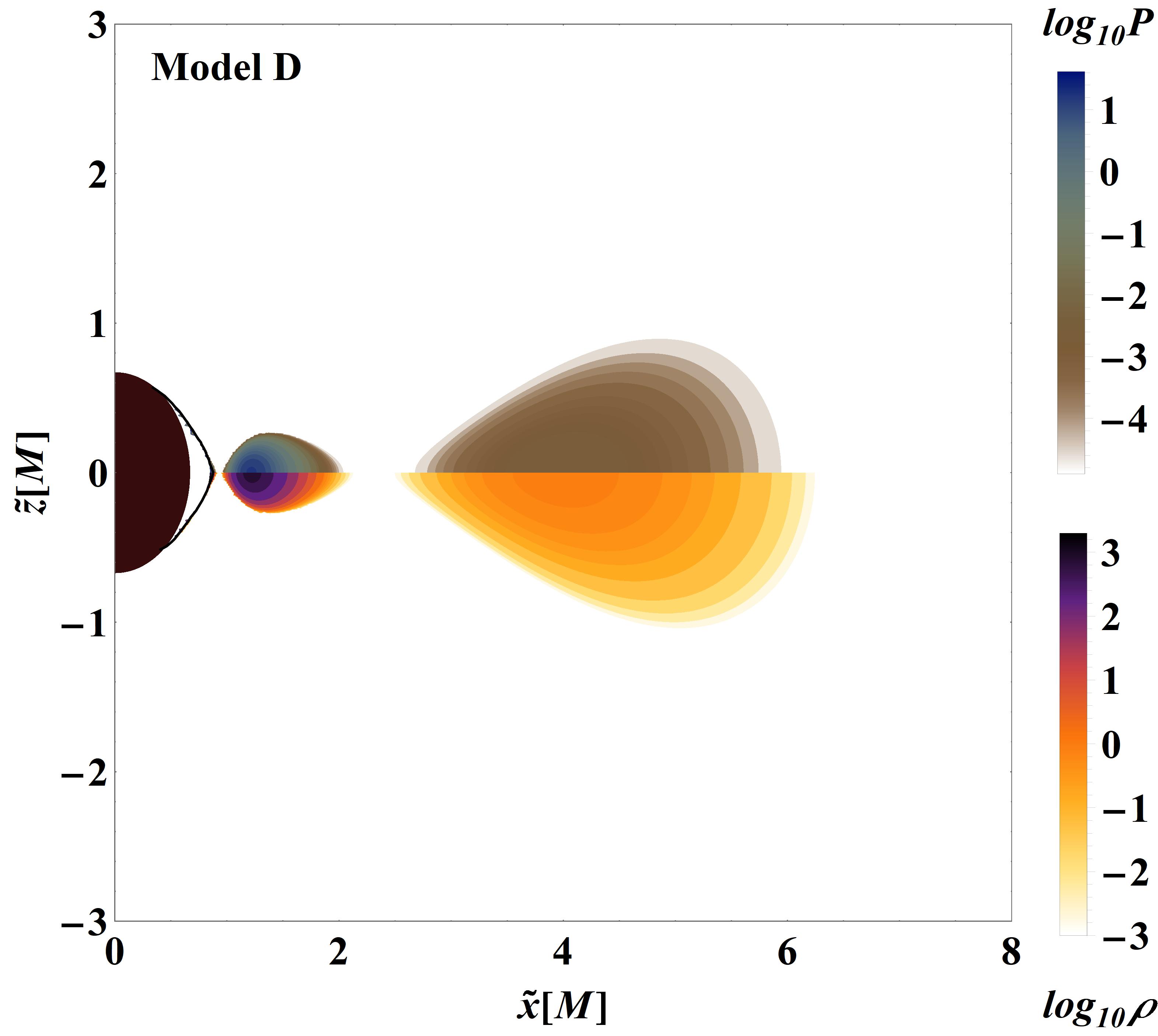}\;\includegraphics[width=5.2cm ]{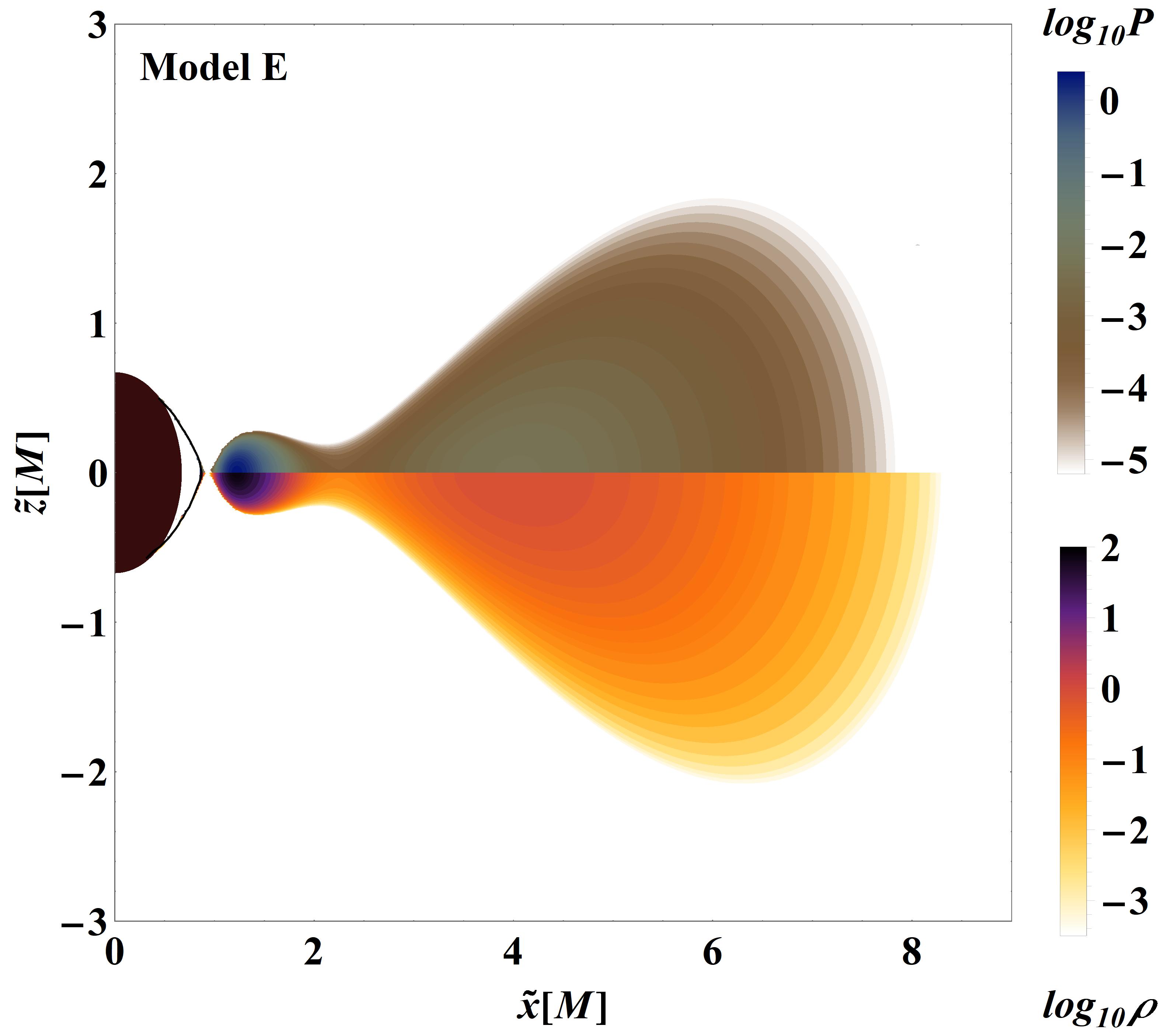}
\caption{Distributions of mass density and pressure in equilibrium tori for Models A-E. In each panel, the upper half of the panel is for the fluid pressure $P$ and the lower half of the panel is for the rest-mass density $\rho$. }\label{fade}
\end{figure}
In Fig.\ref{fade}, we also present the distributions of mass density and pressure in equilibrium tori by solving Eq.(\ref{wwinp}) with the equation of state $P=K w^{\kappa}$. Here, we set $\kappa=4/3$ and $w_{r_{max}}=1$, and then the coefficient $K$ can be calculated by $K=W_{r_{max}}-W_{in}$ for different models, where $W_{r_{max}}$ is the value of $W_{eff}$ at the centre $r=r_{max}$ of the equilibrium torus. Comparing Fig. \ref{fade} with Figs. \ref{f5} and \ref{f51}, one can find that the distributions of mass density and pressure in equilibrium tori are similar to the distribution of the effective potential $W_{eff}$.

\begin{figure}[ht]
\includegraphics[width=5.4cm ]{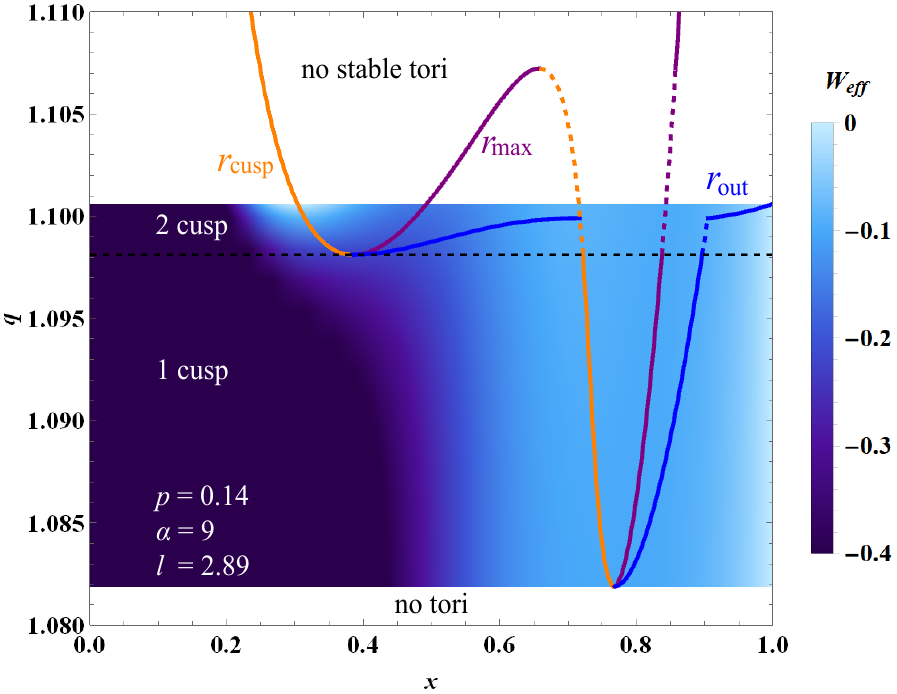}\;\includegraphics[width=5.3cm ]{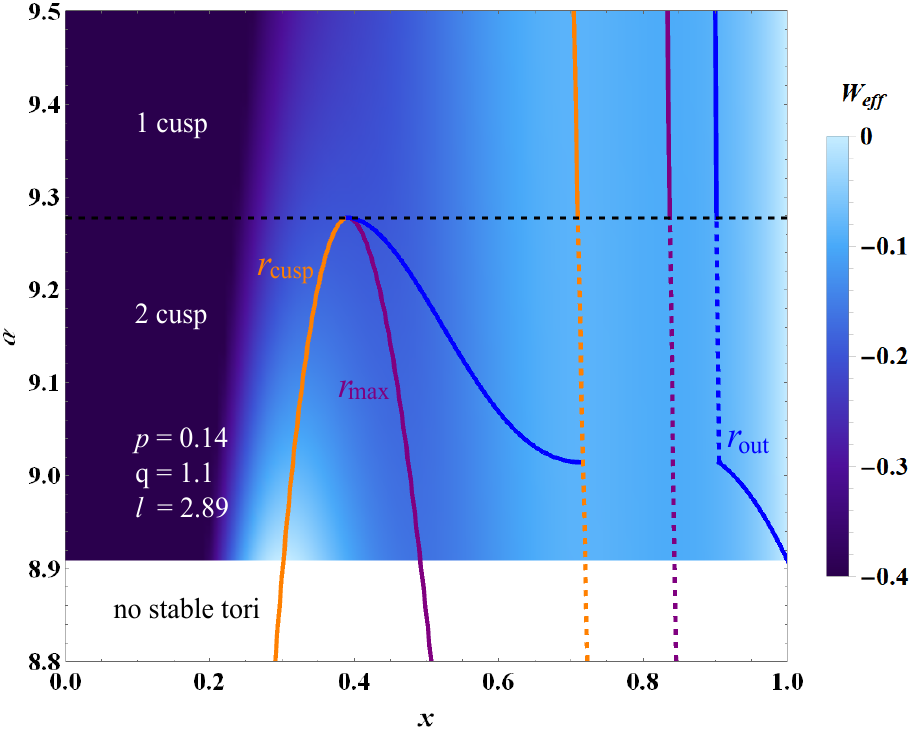}\;\includegraphics[width=5.4cm ]{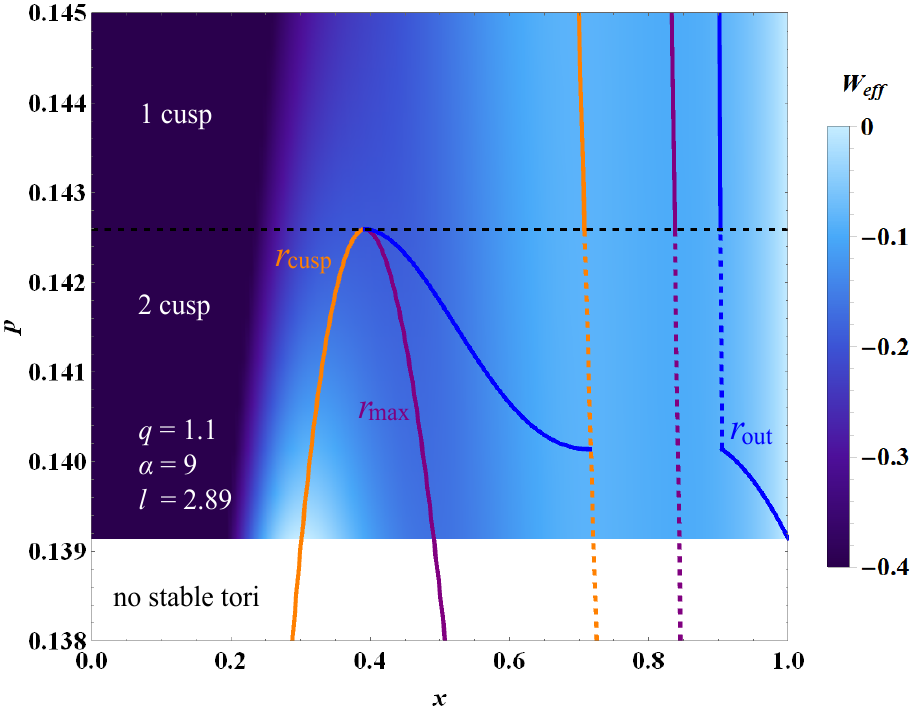}
\caption{Colourmap of the effective potential $W_{eff}$ shown as a function of the conformal radial coordinate $x$ and parameters $q$, $p$ and $\alpha$ for the dyonic black hole with the fixed specific angular momentum $l=2.89$. White regions refer to situations in which $l>l_{mb}$ (no stable tori) or  $l<l^{-}_{ms}$ (no tori) and any equilibrium torus can not be built in these regions.  }\label{f4}
\end{figure}
Fig.\ref{f4} presents the whole set of possible tori solutions appeared in the dyonic black hole  spacetime (\ref{metric0}) for fixed $l=2.89$ and different $p$, $q$ and $\alpha$. In each panel, three different lines describe the most important properties of the dyonic black hole with torus. The orange and purple lines respectively are for the radial positions of the cusp $r_{cusp}$ and the torus centre $r_{max}$. The blue line is for the outer edge of the torus $r_{out}$. It is shown that for fixed $p=0.14$ and $\alpha=9$, we find that in the range $q>1.1006$ (no stable tori $l>l_{mb}$ ) or  $q<1.0819$ (no tori  $l<l^{-}_{ms}$), there is no equilibrium torus around the black hole. There is double tori as $1.0981<q<1.1006$ and only a single torus as $1.0819<q<1.0981$. For fixed $p=0.14$ and $q=1.1$, the double equilibrium tori exist only if $8.9086<\alpha<9.2774$ and the single torus exists as $9.2774<\alpha<151.14$.
For fixed $q=1.1$ and $\alpha=9$, the range of $p$ where double equilibrium tori exist is $0.13914<p<0.14259$, but as $0.14259<p<0.39124$, there is only a single torus in this case.

 Finally, we briefly discuss the possible connection between the parameters for double tori in the dyonic black hole spacetime (\ref{metric0}) and in the parametric Rezzolla-Zhidenko spacetime studied by Cassing \textit{et. al} \cite{2015MNRAS52222M}. Making use of the compactified radial coordinate $x=1-r_H/r$, one can find that the metric ansatz for the dyonic black hole spacetime (\ref{metric0}) can be expressed as
 \begin{equation}
  f(x)=x A(x),\;\;\;\;\;\;\;\;\;\;\;\;B(x)=1,
 \end{equation}
where $A(x)$ can be expanded as a similar form in \cite{Rezzolla:2014mua,2015MNRAS52222M}, i.e.,
 \begin{equation}
  A(x)=1-\epsilon(1-x)+(a_0+\epsilon)(1-x)^2+a_1(1-x)^3+...,
 \end{equation}
 where
 \begin{equation}
  \epsilon=1-\frac{2M}{r_H},\quad\quad a_0=\frac{p^2+q^2}{r^2_H},\quad\quad a_1=1+\frac{p^2+q^2}{r^2_H}-\frac{2M}{r_H}.
 \end{equation}
 Comparing with the Rezzolla-Zhidenko black hole case, it is easy to find that the parameter $\epsilon$ has the same form. The Rezzolla-Zhidenko expansion coefficients $\epsilon$, $a_0$ and $a_1$ are the functions of $q$, $p$ and $\alpha$ because the event horizon $r_H$ also depends on $\alpha$ for the dyonic black hole (\ref{metric0}). Moreover, we find that coefficients $\epsilon$, $a_0$ and $a_1$ are related by $a_1=a_0+\epsilon$ in this case. Thus, these results could help to understand  equilibrium tori around the dyonic black hole and its thick accretion disk.

\section{Summary}

It is useful to study non-self gravitating equilibrium tori around
black holes because they have been widely applied to numerical simulation of accretion flows into black
holes as a kind of initial conditions. We study geometrically thick non-self gravitating equilibrium tori orbiting a static spherically symmetric dyonic black hole. Within the allowed space of parameters, we find that there exist  standard single torus solutions and non-standard double tori solutions. For the single torus solutions, the properties of the equilibrium torus including the cusp and the torus centre are very similar to
those in the Schwarzschild black hole spacetime. For the double tori solutions, the properties of equilibrium tori and the accretion near black hole become far richer. When the value of the potential $W_{eff}$ at the inner cusp is less than one at the outer cusp, the fluid particle with $W_{eff}(r^{-}_{cusp})<W_{eff}<W_{eff}(r^{+}_{cusp})$ can be accreted into the black hole as it is inside the inner torus even if its potential is less than one at the outer cusp. However, if the fluid particle is located inside the outer torus, it can still keep equilibrium  although its equipotential is more than one at the inner cusp. When the value of the potential $W_{eff}$ at the inner cusp is larger than one at the outer cusp, the fluid particles with $W_{eff}(r^{+}_{cusp})<W_{eff}<W_{eff}(r^{-}_{cusp})$ can move along the circular orbit and cannot be accreted into the black hole.  Thus, in the double equilibrium tori case, the occurrence of fluid particle accretion depends on the potential values at the two cusps and the position of fluid particle.  More interestingly, the transitions between single torus and double tori
solutions may occur by changing the specific angular momentum of the fluid. Thus, the electric and magnetic charges and the coupling parameter in the quasitopological electromagnetic theory lead to  a much richer class
of equilibrium tori than in the usual static spherically symmetric cases. The possible
observable effects originating from the equilibrium tori around the dyonic black hole could provide unique tests of general
relativity.

\section{\bf Acknowledgments}

This work was  supported by the National Natural Science
Foundation of China under Grant No.12275078, 11875026, 12035005, and 2020YFC2201400.

\end{document}